\documentclass[12pt,a4paper]{article}
\usepackage{amssymb,amsmath,euscript}
\usepackage{axodraw}

\makeatletter
\@addtoreset{equation}{section}
\makeatother

\newcounter{multieqs}

\newcommand{\be}{\begin{equation}}
\newcommand{\ee}{\end{equation}}

\def\bea{\begin{eqnarray}}
\def\eea{\end{eqnarray}}

\def\Li{{$\rm Li}_2$}

\def\beqa{\begin{eqnarray}}
\def\eeqa{\end{eqnarray}}
\def\beq{\begin{equation}}
\def\eeq{\end{equation}}

\def\one{\mbox{1 \kern-.59em {\rm l}}}

\def\lt{\tilde{\lambda}}

\def\uno{\mbox{1 \kern-.59em {\rm l}}}

\def\lan{\langle}
\def\ran{\rangle}

\def\one{1\!\!1\,\,}

\newcommand{\tr}{\mbox{tr}}

\def\Box{\square}

\def\bcomment#1{}

\def\Li{{\rm Li}_2}
\def\Li2{{\rm Li}_2}

\def\Abar{{\bar A}}
\def\Bbar{{\bar B}}

\begin{document}



\newcommand{\Acal}{{\mathcal A}}
\newcommand{\Rcal}{{\mathcal R}}
\newcommand{\Dcal}{{\mathcal D}}
\newcommand{\Mcal}{{\mathcal M}}
\newcommand{\Ncal}{{\mathcal N}}
\newcommand{\Lcal}{{\mathcal L}}
\newcommand{\Scal}{{\mathcal S}}
\newcommand{\Wcal}{{\mathcal W}}
\newcommand{\Bcal}{\mathcal{B}}
\newcommand{\Ccal}{\mathcal{C}}
\newcommand{\Vcal}{\mathcal{V}}
\newcommand{\Ocal}{\mathcal{O}}
\newcommand{\Qcal}{\mathcal{Q}}


\newcommand{\Urm}{{\mathrm U}}
\newcommand{\Srm}{{\mathrm S}}
\newcommand{\SO}{\mathrm{SO}}
\newcommand{\Sp}{\mathrm{Sp}}
\newcommand{\SU}{\mathrm{SU}}
\newcommand{\Zset}{{\mathbb Z}}
\newcommand{\Cset}{{\,\,{{{^{_{\pmb{\mid}}}}\kern-.47em{\mathrm C}}}}}


\newcommand{\zb}{{\bar{z}}}
\newcommand{\Zb}{\overline{Z}}
\newcommand{\tQ}{\tilde{Q}}
\newcommand{\trho}{\tilde{\rho}}
\newcommand{\tphi}{\tilde{\phi}}
\newcommand{\tlambda}{\tilde{\lambda}}
\newcommand{\dagphi}{{\phi^\dagger}}
\newcommand{\dagq}{{q^\dagger}}
\newcommand{\dagz}{{z^\dagger}}
\newcommand{\bzeta}{{\bar{\zeta}}}
\newcommand{\blambda}{{\bar{\lambda}}}
\newcommand{\bchi}{{\bar{\chi}}}
\newcommand{\tmu}{\tilde{\mu}}
\newcommand{\bA}{\bar{A}}
\newcommand{\bB}{\bar{B}}
\newcommand{\ba}{\bar{a}}
\newcommand{\bq}{\bar{q}}
\newcommand{\bp}{\bar{p}}


\newcommand{\doublet}[2]{\left(\begin{array}{c}#1\\#2\end{array}\right)}
\newcommand{\twobytwo}[4]{\left(\begin{array}{cc} #1&#2\\#3&#4\end{array}\right)}
\newcommand{\ip}[1]{\langle #1\rangle}


\newcommand{\p}{\partial}
\newcommand{\Dslash}{\not{\hbox{\kern-4pt $D$}}}
\newcommand{\half}{\frac{1}{2}}
\newcommand{\diff}{\mathrm{d}}
\newcommand{\ra}{\rightarrow}
\newcommand{\vp}{{\vec{p\,}}}
\newcommand{\dslash}{\slash{\mathrm d}}
\newcommand{\dbar}{\bar{\partial}}
\newcommand{\sint}{{\textstyle \int}}
\newcommand{\Supertwistor}{\mathrm{C} \mathrm{P}^{3|4}}
\newcommand{\Twistorspace}{\mathrm{C} \mathrm{P}^{3}}
\newcommand{\pplfour}{{p_+^1p_+^2p_+^3p_+^4}}
\newcommand{\comment}[1]{{}}
\newcommand{\note}[2]{{\footnotesize [#1}---{\footnotesize \sc  #2]}}
\newcommand{\Yrm}{{\mathrm{Y}}}

\begin{flushright}
QMUL-PH-07-09
\end{flushright}

\vspace{20pt}

\begin{center}

{\Large \bf  One-loop MHV Rules and Pure Yang-Mills }\\
\vspace{33pt}

{\bf {\mbox{Andreas  Brandhuber,  Bill Spence, Gabriele  Travaglini and Konstantinos Zoubos}}}%
\footnote{{\{{\tt a.brandhuber, w.j.spence, g.travaglini, k.zoubos}\}{\tt @qmul.ac.uk }}}

{\em Centre for Research in String Theory \\ Department of Physics\\
Queen Mary, University of
London\\
Mile End Road, London, E1 4NS\\
United Kingdom}
\vspace{40pt}

{\bf Abstract}

\end{center}

\noindent
It has been known for some time that the standard MHV diagram formulation of perturbative Yang-Mills
theory is incomplete,
as it misses rational terms in one-loop scattering amplitudes of pure Yang-Mills.
We propose that certain Lorentz violating counterterms, when expressed in the field variables which
give rise to standard MHV vertices,
produce precisely these missing terms. These  counterterms appear when
Yang-Mills is treated with a regulator, introduced by Thorn and collaborators,  which
arises in worldsheet formulations of Yang-Mills theory in the lightcone gauge.
As an illustration of our proposal, we show that a simple one-loop, two-point counterterm is
the generating function for the infinite sequence of one-loop, all-plus
helicity amplitudes in pure Yang-Mills, in complete agreement with known expressions.

\vspace{0.5cm}

\setcounter{page}{0}
\thispagestyle{empty}
\newpage

\tableofcontents


\setcounter{footnote}{0}

\section{Introduction}

One of the success stories arising from twistor string theory \cite{witten}
(see \cite{Cachazo:2005ga} for a review)
has been the development of new techniques
in perturbative  quantum field theory.
These include recursion relations \cite{bcf,rec},
generalised unitarity \cite{bcfgenun}
and MHV methods (see \cite{BT06} for a review).
One of the key motivations of this work is to provide new approaches
to study and derive
phenomenologically relevant scattering amplitudes.
In particular, this requires that one be able to deal with non-supersymmetric
theories, and to include fermions, scalars, and particles with masses.
A vital first step is to apply these new methods to pure Yang-Mills (YM) theory,
and indeed, some of the first new results inspired by twistor string theory
involved pure YM amplitudes at tree-
\cite{csw,gv,zhu,wu-zhu1,wu-zhu2,gnv,lnv,Bern:2004ba} and
one-loop \cite{bbst2} level.

A recalcitrant issue in this work is the derivation of rational terms in quantum amplitudes.
Unitarity-based techniques \cite{bddk} and loop MHV methods \cite{bst04} 
are successful in obtaining the cut-constructible parts of amplitudes; essentially this is because
at some level they are dealing with four-dimensional cuts. In principle
performing $D$-dimensional cuts generates all parts of amplitudes
\cite{vanNeerven:1985xr,Bern:1995db,Bern:1996ja,bmst} as long as only massless
particles are involved,
however these techniques still appear to be relatively cumbersome.
Combinations of recursive techniques and unitarity have led to important progress
recently \cite{b1,b2,b3,z1,z2,z3,Anastasiou:2006jv,pm,Britto:2006fc,Anastasiou:2006gt},
but it would be preferable to have a more powerful prescriptive formulation,
particularly keeping in mind that applications to more general situations
are sought.

A promising development from this point of view is the Lagrangian approach
\cite{Gorsky:2005sf,Mansfield,EttleMorris}.
Here it has been argued that lightcone Yang-Mills theory, combined with
a certain change of field variables, yields a classical action which
comprises precisely the MHV vertices. A full Lagrangian description of
MHV techniques would in principle give a prescription for
applying such methods to diverse theories.
The next step in developing this is to understand the quantum corrections in this
Lagrangian approach.
If one directly uses  in  a path integral the classical MHV action, containing only purely
four-dimensional MHV vertices,  then it is
immediately clear that this cannot yield all known quantum amplitudes. For example,
there is  no way to construct one-loop amplitudes where the external gluons
all have positive helicities,
or where only one gluon has negative helicity, as all MHV vertices contain
two negative helicity particles
(this issue has been recently discussed in  \cite{BST06}).
These amplitudes are particular cases where the entire amplitude consists of
rational terms. More generally, it seems clear
 that the vertices of the classical MHV Lagrangian will
not yield the rational parts of amplitudes, but only the
cut-constructible terms \cite{bbst2}.
Important insights into this question can be obtained from the study of
self-dual Yang-Mills theory, which
has the same all-plus one-loop amplitude of full YM \cite{Cangemi1,Cangemi2,ChalmersSiegel96}
as its sole quantum correction.%
\footnote{In real Minkowski space, this is in fact
its single non-vanishing amplitude.}
An example, relevant to
the discussion in this paper, is given in
\cite{BST06} where it was shown how these amplitudes might be obtained
from the Jacobian arising from a B\"{a}cklund-type
change of variables which takes the self-dual
Yang-Mills theory to a free theory.

A discussion of the full Yang-Mills theory in the lightcone gauge
has recently been given by Chakrabarti, Qiu and Thorn (CQT) in  \cite{Thorn05,CQT1,CQT2}.
These papers employ an interesting regularisation which,
importantly, does not change the dimension of spacetime.
For this reason, we  find it  particularly suitable for setting the scene for the
MHV diagram method,
which is inherently four-dimensional in current approaches.
The regularisation of CQT   requires the introduction of
certain counterterms, which prove to be rather
simple in form. %
What we will show in this paper is that these simple counterterms
provide a very compact and powerful way to represent the rational terms
in gauge theory amplitudes; specifically,
we  will demonstrate  that the single two-point counterterm
contains {\it all} the
$n$-point all-plus amplitudes. The way this happens is through the use of
the new field variables of \cite{Gorsky:2005sf,Mansfield,EttleMorris}.
Other counterterms will combine with
vertices from the Lagrangian and should generate the rational parts of more general
amplitudes. Based on the discussion in this paper, we  propose that the
counterterms, expressed in the field variables which give rise
to standard MHV vertices, in combination
with Lagrangian vertices, generate the
rational terms previously missing from MHV diagram formulations.

The rest of the paper is organised as follows.
After giving some background material in section 2,
we explicitly derive in section 3 the
four point all-plus amplitude from the two-point counterterm of
CQT. We follow this
by showing that the $n$-point expression,  obtained by writing the counterterm
in new variables, has precisely the right collinear and soft limits required
for it to be the correct all-plus $n$-point amplitude.
We present our conclusions in section 4, and our notation and derivations
of certain identities have been collected in two appendices.

\section{Background}

In this section, we first review the classical field redefinition from the 
lightcone Yang--Mills Lagrangian to the MHV--rules Lagrangian. We then move
on to motivate the four--dimensional regularisation scheme we will employ,
and argue that it leads directly to the introduction of a certain Lorentz--violating
counterterm in the Yang--Mills Lagrangian.  We close the section with the
remarkable observation that this counterterm  provides a simple way to calculate 
the four--point all-plus one--loop amplitude using only tree--level combinatorics.

\subsection{The classical MHV Lagrangian}

It seemed clear from the beginning that the
MHV diagram approach to
Yang-Mills  theory must be closely related to lightcone gauge theory.
This idea was substantiated by Mansfield \cite{Mansfield} (see also \cite{Gorsky:2005sf}).
The starting point of \cite{Mansfield} is the lightcone gauge-fixed YM Lagrangian for
the fields corresponding to the two physical polarisations of the gluon.
It was argued convincingly in \cite{Mansfield} that a certain  canonical change of the field
variables re-expresses this
lightcone Lagrangian as a theory containing the infinite series of MHV vertices.
Some of the arguments in \cite{Mansfield} were rather general; these were reviewed
in \cite{EttleMorris}, where the change of variables was discussed in more
detail, and in particular it was shown how the four- and
five-point MHV vertices arise from the change of variables.
In this paper we will mainly follow the notation of \cite{EttleMorris}.

The general structure of the lightcone YM Lagrangian, after integrating out unphysical degrees
of freedom, is (see appendix \ref{notation} for more details)
\be \label{lcgeneral}
\Lcal_{\rm YM}=\Lcal_{+-}+\Lcal_{++-}+\Lcal_{--+}+\Lcal_{++--} \ ,
\ee
where the gauge condition is $\eta^\mu A_\mu = 0$ with the null vector
$\eta = (1/\sqrt{2},0,0,1/\sqrt{2})$.
Since this Lagrangian contains a $++-$ vertex, it is not of MHV
type. In \cite{Mansfield}, Mansfield proposed to eliminate this
vertex through a suitably chosen field redefinition. Specifically,
he performed a canonical change of variables from $(A,\bar A)$
to new fields $(B,\bar B)$, in such a way that
\be\label{mansf}
\Lcal_{+-}(A,\Abar) +  \Lcal_{++-}(A,\Abar) = \Lcal_{+-}(B,\Bbar)
\ .
\ee
The remarkable result is that upon
inserting this change of variables into the remaining two
vertices, the Lagrangian, written in terms of $(B,\Bbar)$,
becomes a sum of MHV vertices,
\be
\label{MHVlag}
\Lcal_{\rm YM}=\Lcal_{+-}+\Lcal_{+--}+\Lcal_{++--}+\Lcal_{+++--}+
\dots
\ .
\ee
The crucial property of Mansfield's transformation
that makes this possible is that, while both $A$ and $\bA$ are
series expansions in the new $B$ fields, $A$ has no dependence on
the $\bB$ fields while $\bA$ turns out to be \emph{linear} in
$\bB$. Thus, since the remaining vertices are quadratic in
the $\bB$, the new interaction vertices have  the helicity configuration
of an MHV amplitude.
Mansfield was also able to show that the explicit form of the vertices
coincides with the CSW off-shell continuation of the Parke-Taylor
formula for the MHV scattering amplitudes, as proposed by
\cite{csw}.

One of the main results of \cite{EttleMorris} was the derivation
of an explicit, closed formula for the expansion of the original fields $(A,\Abar)$
in terms of the new fields $(B,\Bbar)$.
This was then used to verify that the new vertices
are indeed precisely the MHV vertices of \cite{csw},
at least up to the five-point level.
We will now briefly review these results.
First, recall that the  positive helicity field $A$ is a function
of the positive helicity $B$ field only. It is expanded as follows:
\be \label{Aexpansion}
A(\vp)=\sum_{n=1}^{\infty}\int_{\Sigma}
\prod_{i=1}^n\frac{\diff^3p^i}{(2\pi)^3} \;
\Delta(\vp,\vp^1,\dots\vp^n) \;
\Yrm(\vp;1\cdots n)\ B(\vp^1)B(\vp^2)\cdots B(\vp^n)
\ ,
\ee
where
$\Delta(\vp,\vp^1,\dots\vp^n) := (2\pi)^3\delta^{(3)}(\vp-\vp^1-\cdots -\vp^n)$. Note
that the $x^-$ coordinate is common to all the fields, which is why we have restricted
the transformation to the lightcone quantisation surface $\Sigma$. 

By inserting this expansion into
\eqref{mansf} and using the requirement that the transformation be canonical,
Ettle and Morris succeeded in deriving a very simple expression
for the coefficients $\Yrm$. After translating to our conventions (see
appendix \ref{notation}), they are given by:
\be
\Yrm(\vp;12\cdots n)=(\sqrt2ig)^{n-1}\frac{p_+}{\sqrt{p_+^1p_+^n}}\
\frac{1}{\ip{12}\ip{23}\cdots\ip{n-1,n}}
\ .
\ee
The first few terms in (\ref{Aexpansion}) are then:
\be
\begin{split}
A(\vp)=&B(\vp)+\sqrt2igp_+
\int_{\Sigma}
\frac{\diff^3p^1\diff^3p^2}{(2\pi)^3}\frac{\delta^{(3)}(\vp-\vp^1-\vp^2)}
{\sqrt{p_+^1p_+^2}}
\frac{1}{\ip{12}}\ B(\vp^1)B(\vp^2) \\
&     -2g^2p_+\int_{\Sigma}
\frac{\diff^3p^1\diff^3p^2\diff^3p^3}{(2\pi)^6}
\frac{\delta^{(3)}(\vp-\vp^1-\vp^2-\vp^3)}
{\sqrt{p_+^1p_+^3}}\frac{1}{\ip{12}\ip{23}}\
B(\vp^1)B(\vp^2)B(\vp^3)\\
&+\cdots
\ .
\end{split}
\ee
\normalsize
Similarly, one can write down the expansion of the negative helicity
field $\bar{A}$, which, as discussed above, is linear
in $\bB$, but is an infinite series in the new field $B$.
In \cite{EttleMorris} it was shown that
the coefficients in the expansion of $\bA$ are very closely related to those for
$A$.\footnote{This is perhaps easiest to see \cite{FengHuang06} by considering 
that, in the context of $\Ncal=4$ SYM, $A$ and $B$ are part of the same lightcone 
superfield.} The expansion of $\bB$ turns out to be simply
%
\be
\begin{split}
\bar{A}(\vp)=&\!-\!\sum_{n=1}^{\infty}\sum_{s=1}^{n}
\int_{\Sigma}
\prod_{i=1}^n\frac{\diff^3p^i}{(2\pi)^3} \;
\Delta(\vp,\vp^1,\dots,\vp^n)
\frac{(p_+^s)^2}{(p_+)^2}\ \Yrm(\vp;1\cdots n)\
B(\vp^1){\cdots} \bar{B}(\vp_s){\cdots} B(\vp^n)\\
&=-\sum_{n=1}^{\infty}
\int_{\Sigma}
\prod_{i=1}^n\frac{\diff^3p^i}{(2\pi)^3} \;
\Delta(\vp,\vp^1,\ldots,\vp^n)\;
\frac{1}{(p_+)^2}\Yrm(\vp;1\cdots n) \\
&\qquad\qquad\times\sum_{s=1}^{n} (p_+^s)^2 \ B(\vp^1)\cdots \bar{B}(\vp^s) \cdots B(\vp^n)
\ .
\end{split}
\ee
\normalsize
Thus we see that at each order in the expansion, we need to sum over all possible
positions of $\bB$. Explicitly, the first few terms are:
\be
\begin{split}
\bar{A}(\vp)&=\bB(\vp)
-\sqrt2ig\int_{\Sigma}
\frac{\diff^3p^1\diff^3p^2}{(2\pi)^3}\delta^{(3)}(\vp-\vp^1-\vp^2)
\frac{1}{p_+\sqrt{p_+^1p_+^2}}\frac{1}{\ip{12}}\times \\
&\qquad\qquad\qquad\qquad
\times\left[(p_+^1)^2\bB(\vp^1) B(\vp^2)+(p_+^2)^2 B(\vp^1)\bB(\vp^2)\right]\\
&+2g^2\int_{\Sigma}
\frac{\diff^3p^1\diff^3p^2\diff^3p^3}{(2\pi)^6}
\delta^{(3)}(\vp-\vp^1-\vp^2-\vp^3)
\frac{1}{p_+\sqrt{p_+^1p_+^3}}\frac{1}{\ip{12}\ip{23}}\times \\
&\times\!\left[(p_+^1)^2 \bB(\vp^1) B(\vp^2) B(\vp^3)\!+\!
(p_+^2)^2B(\vp^1) \bB(\vp^2) B(\vp^3)
\!+\!(p_+^3)^2B(\vp^1)B(\vp^2)\bB(\vp^3)
\right]\\
&+\cdots
\end{split}
\ee
\normalsize
Using the above results, it is  in principle  straightforward
 to derive the terms that arise on inserting the Mansfield
transformation into the two remaining vertices of the theory.
For the simplest cases, one can see explicitly that these combine
to produce MHV vertices, and some arguments were also given
in \cite{Mansfield,EttleMorris} that this must be true in general.

In supersymmetric theories, the MHV vertices are enough to reproduce
complete scattering amplitudes at one loop \cite{BST05}.
However, as we mentioned earlier,
for pure YM it is clear that the terms in the MHV Lagrangian
\eqref{MHVlag} will not be enough to generate complete  quantum amplitudes.
For instance, the scattering amplitude with all gluons
with positive helicity, which at one loop is finite and given by a rational term,
cannot be obtained by only using MHV diagrams, for the simple reason that one
cannot draw any diagram contributing to it by only resorting to
MHV vertices.%
\footnote{On the other hand, it was shown in \cite{BST06} that the parity conjugate
all-minus amplitude is correctly  generated by using MHV diagrams.}
Another amplitude which cannot be derived within conventional
MHV diagrams is the amplitude with only one gluon of negative helicity.
Similarly to the all-plus amplitude, this single-minus amplitude
vanishes at tree level, and at one loop is given by a finite, rational
function of the spinor variables.

The lesson we learn from this is that,
in order to apply the MHV method to derive complete amplitudes in pure YM, one should
look more closely at the change of variables in the full quantum theory.
 There are several possible subtleties one should pay careful attention
to at the quantum level. First of all, it is possible that the canonical
nature of the transformation is not preserved,
leading to a non--trivial Jacobian  which could  provide
the missing amplitudes. Another possible source of contributions could come from
violations of the equivalence theorem. This theorem states that,
although correlation functions of the new fields are in general different
from those of the old fields, the scattering amplitudes are actually the same%
\footnote{Modulo a trivial wave-function renormalisation.},
as long as the new fields are good interpolating fields.
These issues were explored in some detail in \cite{BST06} (see also
\cite{EttleMorris,FengHuang06}) where it was shown, for a different (non-canonical)
field redefinition, how a careful treatment of these effects can combine to
reproduce some of the amplitudes that would seem to be missing at first sight.

Another aim of \cite{BST06} was to demonstrate how to reproduce one of the
above--mentioned rational amplitudes, the one with all--minus helicities,
in the MHV formalism. This amplitude is slightly less mysterious than
the all--plus amplitude in the sense that one can write down the
contributing diagrams using only MHV vertices; however a  calculation without
a suitable regulator in place would give a vanishing answer, 
despite the fact that this amplitude is finite. In \cite{BST06},
it was shown, using dimensional regularisation,
that the full nonzero result arises
from a slight mismatch between four-- and $D$ ($=4-2\epsilon$)--dimensional momenta.

 It is natural therefore to expect that dimensional regularisation will be helpful
also for the problem at hand, which is to recover the rational amplitudes of
pure Yang--Mills after the Mansfield transformation. Decomposing the regularised
lightcone Lagrangian into a pure four-dimensional part and the remaining 
$\epsilon$--dependent terms, and performing the transformation on the 
four-dimensional part only, will give rise to several new $\epsilon$--dependent terms 
that can potentially give finite answers when forming loops.

Although this approach shows promise, it is not the one we will make use of
in the following. Instead, motivated by the fact that the Mansfield transformation
seems to be deeply rooted in four dimensions, we would like to look for
a purely four--dimensional regularisation scheme. We now turn to a review
of the particular scheme we will use.

\subsection{A four--dimensional regulator for lightcone Yang--Mills}

 In the above we explained why a na\"{\i}ve application of the
Mansfield transform leads to puzzles at the quantum level, and
discussed possible ways to improve the situation. The conclusion
was that, since the missing amplitudes arise from subtle mismatches in
regularisation, one should be careful to perform the Mansfield transform
on a suitably regularised version of the lightcone Yang--Mills action.
Here we will review one approach to the regularisation of lightcone
Yang--Mills, which, despite several slightly unusual features, appears to
be ideally suited for the problem at hand.

The regularisation we propose to use is inspired by recent work
of CQT \cite{Thorn05,CQT1,CQT2}
on Yang--Mills amplitudes in the lightcone worldsheet approach \cite{BT01,Thorn02}.
This is an attempt to understand gauge--string duality
which is similar in spirit to 't Hooft's original work on the planar limit of
gauge theory \cite{tHooft74}, and aims at improving on early dual model techniques
\cite{NielsenOlesen70,SakitaVirasoro70}. We recall that one of the main goals in
those works is to exhibit the string worldsheet as made up of very large planar
 diagrams (``fishnets'').

In their recent work, Thorn and collaborators make this statement more
precise, using techniques that were unavailable when the original ideas were put
forward. It is hoped that, by understanding how to translate a generic Yang--Mills
planar diagram to a configuration of fields (with suitable boundary conditions)
on the lightcone worldsheet, it will eventually become possible to perform the
sum of all these diagrams. This approach to gauge--string duality is thus
complementary to that using the AdS/CFT correspondence.

The field content and structure of the worldsheet theory dual to Yang--Mills
theory is rather intricate (see e.g.~\cite{Thorn02}), but for our purposes the details
are not important. What is most relevant for us is that one of the principles
of this approach is that all quantities on the Yang--Mills side should have
a local worldsheet description. This includes the choice of regulator that needs
to be introduced when calculating loop diagrams. This requirement led Thorn
\cite{Thorn04} (see also \cite{BT02a,Bardakci03}) to introduce an exponential UV cutoff, 
which we will discuss in a short while.

  Since one of the goals of this programme is to translate an arbitrary planar
diagram into worldsheet form (and eventually calculate it), it is an important
intermediate goal to understand how to do standard Yang--Mills perturbation theory
in ``worldsheet--friendly" language. In \cite{Thorn05,CQT1,CQT2}
CQT do exactly that for the simplest case, that of one--loop
diagrams in Yang--Mills theory, by analysing how familiar features like renormalisation
are affected by the unusual regularisation procedure and other special features of the
lightcone worldsheet formalism.

To conclude this brief overview of the lightcone worldsheet formalism, the main
point for our current purposes 
is that it provides motivation and justification
for a slightly unusual  regularisation of lightcone Yang--Mills,
which we will now describe.

Let us momentarily focus on the
self--dual part of the lightcone Yang--Mills Lagrangian:
\be
\label{lclaga}
\Lcal=\Lcal_{-+}+\Lcal_{++-}=-A_\zb\Box A_z
+2ig[A_z,\p_+A_\zb](\p_+)^{-1}(\p_\zb A_z)
\ .
\ee
This action provides one of the representations of self-dual Yang-Mills theory. 
After transforming to momentum space, we find that the only
vertex in the theory is the following (suppressing the gauge index structure):%

\be
\begin{picture}(30,30)(5,10)
\put(0,0){
\SetColor{BrickRed}
\Line(5,5)(15,15)
\Line(25,5)(15,15)
\Line(15,15)(15,25)
\Text(4,4)[tr]{$A_2$}
\Text(26,4)[tl]{$A_1$}
\Text(15,27)[bc]{$\bar{A}_3$}
}
\end{picture}
=-2g\frac{p_+^3}{p_+^1p_+^2}[p_+^1p_\zb^2-p_+^2p_\zb^1]=
-\sqrt{2}g\frac{p_+^3}{\sqrt{p_+^1p_+^2}}~[12] \ .
\ee

As for propagators, following \cite{CQT1}, we will use
the Schwinger representation:
\be \label{Schwinger2}
\frac{1}{p^2}=-\int_0^\infty\diff T e^{+Tp^2}
\ .
\ee
In (\ref{Schwinger2}) $p^2$ is understood to be the appropriate ($p^2<0$) 
Wick rotated version of the Minkowski space inner product. For our
choice of signature, the latter is
\be
p\cdot q=p_+q_-+p_-q_+-{\bf p\cdot q}=p_+q_-+p_-q_+-(p_zq_\zb+p_\zb q_z)
\ ,
\ee
so that $p^2=2(p_+p_--p_zp_\zb)$. 

We will also make use of the dual or \lq\lq region momentum\rq\rq\ representation,
where one assigns a momentum to each \emph{region} that is bounded by a line
in the planar diagram. By convention,
the actual momentum of the line is given by the region momentum to its
right minus that on its left, as given by the direction of momentum flow%
\footnote{In \cite{CQT1} the flow of momentum is chosen to always match
the flow of helicity, but we will not use this convention.}.
Clearly such a prescription can only be straightforwardly
implemented for planar diagrams, which is the case considered in \cite{CQT1}.
This is also sufficient for our purposes, since we are calculating the
leading single--trace contribution to one--loop scattering amplitudes.
Non--planar (multi--trace) contributions can
be recovered from suitable permutations of the leading--trace ones
(see e.g.~\cite{BDDK94}).

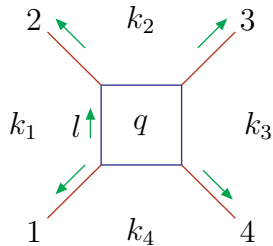
\begin{figure}[h]
\begin{center}
\begin{picture}(100,100)(0,0)
\put(0,0){
\SetColor{BrickRed}
\Line(10,10)(30,30)
\Line(10,80)(30,60)
\Line(80,80)(60,60)
\Line(80,10)(60,30)
\SetColor{Blue}
\Line(30,30)(30,60)
\Line(30,60)(60,60)
\Line(60,60)(60,30)
\Line(60,30)(30,30)
\SetColor{Green}
\LongArrow(24,30)(14,20)
\LongArrow(24,74)(14,84)
\LongArrow(66,74)(76,84)
\LongArrow(68,28)(78,18)
\LongArrow(26,40)(26,50)
\Text(45,45)[c]{$q$}
\Text(1,45)[c]{$k_1$}\Text(45,85)[c]{$k_2$}\Text(89,45)[c]{$k_3$}\Text(45,5)[c]{$k_4$}
\Text(5,5)[c]{$1$}\Text(5,85)[c]{$2$}\Text(85,85)[c]{$3$}\Text(85,5)[c]{$4$}
\Text(21,45)[c]{$l$}
}
\end{picture}
\caption{A sample one--loop diagram indicating the labelling of region momenta.
The outgoing leg momenta
are $p_1=k_1-k_4\;,\;p_2=k_2-k_1\;,\;p_3=k_3-k_2\;,\;p_4=k_4-k_3$, while the loop momentum
(directed as indicated) is $l=q-k_1$.}\label{Sample}

\end{center}
\end{figure}

To demonstrate the use of region momenta, a sample one--loop diagram is
pictured in Figure \ref{Sample}.

The ``worldsheet--friendly'' regulator that CQT employ is simply defined as
follows \cite{Thorn04}:
For a general $n$--loop diagram, with $q_i$ being the loop region momenta,
one simply inserts an exponential cutoff factor
\be\label{regulator}
\mathrm{exp}(-\delta\sum_{i=1}^n{\bf q}_i^2)
\ee
in the loop integrand, where $\delta$ is positive and will be taken to zero at the end
of the calculation.
This clearly regulates UV divergences (from large transverse momenta),
but, as we will see, has some surprising consequences
since it will lead to finite values for certain Lorentz--violating processes,
which therefore have to be cancelled by the introduction of appropriate counterterms.

Note that  ${\bf q}^2=2q_z q_\zb$ has components only along the
two transverse directions, hence it breaks explicitly even more  Lorentz
invariance than the lightcone usually does. This might seem rather unnatural from
a field-theoretical point of view,  however it is crucial in the lightcone worldsheet approach.
Indeed, the lightcone time
$x^-$ and $x^+$ (or in practice its dual momentum $p_+$)
parametrise the worldsheet
itself,  and are regulated by discretisation; thus, they are necessarily treated very
differently from the two transverse momenta $q_z,q_\zb$ which appear as
dynamical worldsheet scalars.
Fundamentally, this is because of the need to preserve longitudinal ($x^+$)
boost invariance (which eventually leads to conservation of discrete $p_+$).
The fact that the regulator depends on the region momenta rather than the actual
ones is a consequence of asking for it to have a local description on the worldsheet.

The main ingredient for what will follow later in this paper is the computation of the $(++)$
one--loop gluon self--energy in the regularisation scheme discussed
earlier. This is performed on page 10 of \cite{CQT1}, and we will briefly outline it here.
This helicity--flipping gluon self--energy, which we denote by $\Pi^{++}$, is the only potential
self--energy contribution in self--dual Yang--Mills; in full YM we would also have $\Pi^{+-}$
and, by parity invariance,  $\Pi^{--}$.

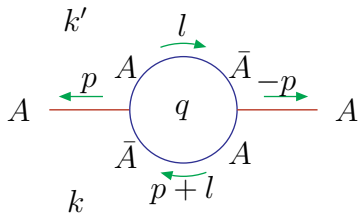
\begin{figure}[h]
\begin{center}
\begin{picture}(100,80)(0,0)
\put(0,0){
\SetColor{BrickRed}
\Line(20,50)(50,50)
\SetColor{Blue}
\CArc(70,50)(20,0,180)
\CArc(70,50)(20,180,0)
\SetColor{BrickRed}
\Line(90,50)(120,50)
\Text(13,50)[r]{$A$}
\Text(127,50)[l]{$A$}
\Text(53,62)[br]{$A$}
\Text(87,62)[bl]{$\bA$}
\Text(87,38)[tl]{$A$}
\Text(53,38)[tr]{$\bA$}
\Text(30,80)[b]{$k'$}
\Text(30,20)[t]{$k$}
\Text(70,50)[c]{$q$}
\SetColor{Green}
\LongArrow(40,55)(25,55)
\Text(35,60)[c]{$p$}
\LongArrow(100,55)(115,55)
\Text(105,60)[c]{$-p$}
\LongArrowArcn(70,50)(25,110,70)
\Text(70,82)[c]{$l$}
\LongArrowArcn(70,50)(25,-70,-110)
\Text(70,20)[c]{$p+l$}
}
\end{picture}
\caption{Labelling of one of the selfenergy diagrams contributing to $\Pi^{++}$.}\label{Bubblefigure}
\end{center}
\end{figure}

There are two contributions to this process, corresponding to the two ways to route helicity
in the loop, but they can be easily shown to be equal so we will concentrate on one of
them, which is pictured in Figure \ref{Bubblefigure}.

In Figure \ref{Bubblefigure},  $p,-p$ are the outgoing line momenta, $l$ is the loop line
momentum, and $k,k',q$ are the region momenta, in terms of which the line momenta
are given by
\be
p=k'-k,\quad l=q-k'
\;.
\ee

Remembering to double the result of this diagram, we find
the following expression for the self--energy:
\be \label{Bubble1}
\begin{split}
\Pi^{++}=&8g^2N\int \frac{\diff^4 l}{(2\pi)^4}\left[
\frac{-(p+l)_+}{p_+l_+}(p_+l_\zb-l_+p_\zb)
\right]\times
\frac{1}{l^2(p+l)^2}\times\\
&\qquad\qquad\quad\times\left[
\frac{-l_+}{(-p_+)(p+l)_+}((-p_+)(p_\zb+l_\zb)-(p_++l_+)(p_\zb))
\right]\\
=&\frac{g^2N}{2\pi^4}\int \diff^4 l\frac{1}{(p_+)^2}(p_+l_\zb-l_+p_\zb)
(p_+(p_\zb+l_\zb)-(p_++l_+)p_\zb)\frac{1}{l^2(p+l)^2}
\ .
\end{split}
\ee
Although we are suppressing the colour structure, the factor of $N$ is
easy to see by thinking of the double--line representation of this diagram\footnote{
For simplicity, we take the gauge group to be $\Urm(N)$.}.
One of the crucial properties of (\ref{Bubble1}) is that the factors of the loop momentum
$l_+$ coming from the vertices have cancelled out, hence there are no potential subtleties
in the loop integration as $l_+\ra0$. This means that, although for general loop
calculations one would have to follow the DLCQ procedure and discretise $l_+$
(as is done for other processes considered in \cite{Thorn05,CQT1,CQT2}), this issue does not
arise at all for this particular integral,  and we are free to keep $l_+$ continuous.

To proceed, we convert momenta to region momenta, rewrite
propagators in Schwinger representation, and regulate divergences using the 
regulator (\ref{regulator}).
Employing the unbroken shift
symmetry in the $+$ region momenta to further set $k_+=0$,  (\ref{Bubble1}) can be recast as:
\be
\begin{split}
\Pi^{++}=&\frac{g^2N}{2\pi^4}\int_0^\infty\diff T_1\diff T_2\int\diff^4q
\frac{1}{(k'_+)^2} e^{T_1(q-k)^2+T_2(q-k')^2-\delta \bf{q}^2}\times\\
&\times \left[k'_+(q_\zb-k'_\zb)-(q_+-k'_+)(k'_\zb-k_\zb)\right]
\left[k'_+(q_\zb-k_\zb)-q_+(k'_\zb-k_\zb)\right]
\ .
\end{split}
\ee
Since $q_-$ only appears in the exponential, the $q_-$ integration will lead to a
delta function containing $q_+$, which can be easily integrated and leads to a
Gaussian--type integral for $q_z,q_\zb$. Performing this integral,  we obtain
(setting $T=T_1+T_2$, $x=T_1/(T_1+T_2)$)
\be
\Pi^{++}=\frac{g^2N}{2\pi^2}\int_0^1\diff x\int_0^\infty\diff T~
\delta^2~\frac{[xk_\zb+(1-x)k'_\zb]^2}{(T+\delta)^3}~
e^{Tx(1-x) p^2-\frac{\delta T}{T+\delta}(x{\bf{k}}+(1-x){\bf k}')^2}
\ .
\ee
Notice that, had we not regularised using the $\delta$ regulator,  we would have
obtained zero at this stage.
Instead, now we can see that there is a region of the $T$ integration
(where $T\sim \delta$) that can lead to a nonzero result.
On performing the $T$ and $x$ integrations,
and sending $\delta$ to zero at the end,
we obtain the following finite answer:

\begin{equation} \label{selfenergy}
\Pi^{++}=2\left(+
\SetScale{0.3}
\begin{picture}(33,15)(0,0)
\put(-5,-12){
\SetColor{BrickRed}
\Line(20,50)(50,50)
\SetColor{Blue}
\CArc(70,50)(20,0,180)
\CArc(70,50)(20,180,0)
\SetColor{BrickRed}
\Line(90,50)(120,50)
}
\end{picture}
+\right)
=\frac{g^2N}{12\pi^2}\left((k_\zb)^2+(k'_\zb)^2+k_\zb k'_\zb\right)
\,.
\ee

Notice that this nonvanishing, finite result violates Lorentz invariance, since it
would imply that a single gluon can flip its helicity. Also, it explicitly depends
on only the $\zb$ components of the region momenta. Such a term is clearly
absent in the tree-level Lagrangian (unlike e.g.~the
$\Pi^{+-}$ contribution in full Yang--Mills theory),
thus it cannot be absorbed through renormalisation --  it will
have to be explicitly cancelled by a counterterm.
This counterterm, which will play a major r\^{o}le in the following,
is defined in such a way that:
\be \label{cteq}
\SetScale{0.3}
\begin{picture}(33,20)(0,0)
\put(-5,-12){
\SetColor{BrickRed}
\Line(20,50)(50,50)
\SetColor{Blue}
\CArc(70,50)(20,0,180)
\CArc(70,50)(20,180,0)
\SetColor{BrickRed}
\Line(90,50)(120,50)
}
\end{picture}
+
\SetScale{0.3}
\begin{picture}(33,20)(0,0)
\put(-5,-12){
\SetColor{BrickRed}
\Line(20,50)(60,50)
\SetColor{Blue}
\CCirc(70,50){10}{Blue}{Green}
\SetColor{BrickRed}
\Line(80,50)(120,50)
}
\end{picture}
=0\ ,
\ee
in other words it will cancel all insertions of $\Pi^{++}$, diagram by diagram.
Let us note here that, had we been doing dimensional regularisation, all bubble
contributions would vanish on their own, so there would be no need to add any
counterterms. So this effect is purely due to the ``worldsheet--friendly''
regulator (\ref{regulator}).

It is also interesting to observe that in a supersymmetric theory this bubble 
contribution would vanish\footnote{This can in fact be derived from the results
of \cite{Qiu06}, where similar calculations were considered with fermions and
scalars in the loop.} so this effect is only of relevance to pure Yang--Mills 
theory.

\subsection{The one--loop (++++) amplitude}

Now let us look at the all--plus four-point one--loop amplitude in this theory.
It is easy to see that it will receive contributions from three types of
geometries: boxes, triangles and bubbles. It is a remarkable property%
\footnote{This
observation is attributed to Zvi Bern \cite{CQT1}.} that the sum of all these
geometries adds up to zero. In particular, with a suitable routing of momenta, the
integrand itself is zero.
Pictorially, we can state this as:
\begin{equation} \label{cancellation}
\begin{picture}(50,50)(0,0)
\put(-10,-25){
\SetColor{BrickRed}
\Line(10,10)(20,20)
\Line(10,50)(20,40)
\SetColor{Blue}
\Line(20,20)(20,40)
\Line(20,40)(40,40)
\Line(40,40)(40,20)
\Line(40,20)(20,20)
\SetColor{BrickRed}
\Line(40,40)(50,50)
\Line(40,20)(50,10)
}
\end{picture}
+4\times
\begin{picture}(60,50)(0,0)
\put(-10,-25){
\SetColor{BrickRed}
\Line(10,10)(20,20)
\Line(10,50)(20,40)
\SetColor{Blue}
\Line(20,20)(20,40)
\Line(20,40)(35,30)
\Line(35,30)(20,20)
\SetColor{BrickRed}
\Line(35,30)(50,30)
\Line(50,30)(60,50)
\Line(50,30)(60,10)
}
\end{picture}
+2\times
\begin{picture}(70,50)(0,0)
\put(-10,-25){
\SetColor{BrickRed}
\Line(10,10)(20,30)
\Line(10,50)(20,30)
\Line(20,30)(30,30)
\SetColor{Blue}
\CArc(40,30)(10,0,360)
\SetColor{BrickRed}
\Line(50,30)(60,30)
\Line(60,30)(70,50)
\Line(60,30)(70,10)
}
\end{picture}
+8\times
\begin{picture}(70,50)(0,0)
\put(-10,-25){
\SetColor{BrickRed}
\Line(10,10)(30,30)
\Line(10,50)(30,30)
\Line(30,30)(50,30)
\Line(50,30)(70,50)
\Line(50,30)(70,10)
\CCirc(20,40){5}{Blue}{White}
}
\end{picture}
=0 \, .
\end{equation}

The coefficients mean that we need to add that number of inequivalent orderings.
So we see (and refer to \cite{CQT1} for the explicit calculation) that the sum of
all the diagrams that one can construct from the single vertex in our theory,
gives a vanishing answer.
However, as discussed in the
previous section, this is not everything: we need to also
include the contribution of the counterterm that we are  forced to add in order to preserve
Lorentz invariance. Since this counterterm, by design,  cancels all the
bubble graph contributions, we are left with just the sum of the box and the
four triangle diagrams.
In pictures,
\be
\Acal^{++++}=
\SetScale{0.6}
\begin{picture}(25,30)(0,0)
\put(-6,-15){
\SetColor{BrickRed}
\Line(10,10)(20,20)
\Line(10,50)(20,40)
\SetColor{Blue}
\Line(20,20)(20,40)
\Line(20,40)(40,40)
\Line(40,40)(40,20)
\Line(40,20)(20,20)
\SetColor{BrickRed}
\Line(40,40)(50,50)
\Line(40,20)(50,10)
}
\end{picture}
+4\times
\begin{picture}(30,30)(0,0)
\put(-6,-15){
\SetColor{BrickRed}
\Line(10,10)(20,20)
\Line(10,50)(20,40)
\SetColor{Blue}
\Line(20,20)(20,40)
\Line(20,40)(35,30)
\Line(35,30)(20,20)
\SetColor{BrickRed}
\Line(35,30)(50,30)
\Line(50,30)(60,50)
\Line(50,30)(60,10)
}
\end{picture}
+\left(2\times
\begin{picture}(35,30)(0,0)
\put(-6,-15){
\SetColor{BrickRed}
\Line(10,10)(20,30)
\Line(10,50)(20,30)
\Line(20,30)(30,30)
\SetColor{Blue}
\CArc(40,30)(10,0,360)
\SetColor{BrickRed}
\Line(50,30)(60,30)
\Line(60,30)(70,50)
\Line(60,30)(70,10)
}
\end{picture}
+8\times
\begin{picture}(35,30)(0,0)
\put(-6,-15){
\SetColor{BrickRed}
\Line(10,10)(30,30)
\Line(10,50)(30,30)
\Line(30,30)(50,30)
\Line(50,30)(70,50)
\Line(50,30)(70,10)
\CCirc(20,40){5}{Blue}{White}
}
\end{picture}
+2\times
\begin{picture}(35,30)(0,0)
\put(-6,-15){
\SetColor{BrickRed}
\Line(10,10)(20,30)
\Line(10,50)(20,30)
\Line(20,30)(60,30)
\CCirc(40,30){3}{Blue}{Green}
\Line(60,30)(70,50)
\Line(60,30)(70,10)
}
\end{picture}
+8\times
\begin{picture}(35,30)(0,0)
\put(-6,-15){
\SetColor{BrickRed}
\Line(10,10)(30,30)
\Line(10,50)(30,30)
\Line(30,30)(50,30)
\Line(50,30)(70,50)
\Line(50,30)(70,10)
\CCirc(20,40){3}{Blue}{Green}
}
\end{picture}
\right)
\ee
where $\Acal^{++++}$ is the known result \cite{beko}
 for the leading--trace part of the four--point all-plus amplitude:
\be \label{allplus-4}
\Acal^{++++}(A_1A_2A_3A_4)=i\frac{g^4N}{48\pi^2}\frac{[12][34]}{\ip{12}\ip{34}}
\ ,
\ee
and the terms in the parentheses clearly cancel among themselves.
This leaves the box and triangle diagrams, which are exactly those appearing in the 
calculation of the parity conjugate amplitude
using dimensional regularisation \cite{BST06}, where the bubbles were zero to begin with.

Following \cite{CQT1}, we make the obvious, but important for the following, observation
that one can change the position of the parentheses:
\be
\Acal^{++++}=
\SetScale{0.6}
\left(
\begin{picture}(30,30)(0,0)
\put(-6,-15){
\SetColor{BrickRed}
\Line(10,10)(20,20)
\Line(10,50)(20,40)
\SetColor{Blue}
\Line(20,20)(20,40)
\Line(20,40)(40,40)
\Line(40,40)(40,20)
\Line(40,20)(20,20)
\SetColor{BrickRed}
\Line(40,40)(50,50)
\Line(40,20)(50,10)
}
\end{picture}
+4\times
\begin{picture}(30,30)(0,0)
\put(-6,-15){
\SetColor{BrickRed}
\Line(10,10)(20,20)
\Line(10,50)(20,40)
\SetColor{Blue}
\Line(20,20)(20,40)
\Line(20,40)(35,30)
\Line(35,30)(20,20)
\SetColor{BrickRed}
\Line(35,30)(50,30)
\Line(50,30)(60,50)
\Line(50,30)(60,10)
}
\end{picture}
+2\times
\begin{picture}(35,30)(0,0)
\put(-6,-15){
\SetColor{BrickRed}
\Line(10,10)(20,30)
\Line(10,50)(20,30)
\Line(20,30)(30,30)
\SetColor{Blue}
\CArc(40,30)(10,0,360)
\SetColor{BrickRed}
\Line(50,30)(60,30)
\Line(60,30)(70,50)
\Line(60,30)(70,10)
}
\end{picture}
+8\times
\begin{picture}(35,30)(0,0)
\put(-6,-15){
\SetColor{BrickRed}
\Line(10,10)(30,30)
\Line(10,50)(30,30)
\Line(30,30)(50,30)
\Line(50,30)(70,50)
\Line(50,30)(70,10)
\CCirc(20,40){5}{Blue}{White}
}
\end{picture}
\right)
+2\times
\begin{picture}(35,30)(0,0)
\put(-6,-15){
\SetColor{BrickRed}
\Line(10,10)(20,30)
\Line(10,50)(20,30)
\Line(20,30)(60,30)
\CCirc(40,30){3}{Blue}{Green}
\Line(60,30)(70,50)
\Line(60,30)(70,10)
}
\end{picture}
+8\times
\begin{picture}(35,30)(0,0)
\put(-6,-15){
\SetColor{BrickRed}
\Line(10,10)(30,30)
\Line(10,50)(30,30)
\Line(30,30)(50,30)
\Line(50,30)(70,50)
\Line(50,30)(70,10)
\CCirc(20,40){3}{Blue}{Green}
}
\end{picture}
\ee
where again the terms in the parentheses are zero (by (\ref{cancellation})). This
demonstrates that one can compute the all-plus amplitude just from a tree-level
calculation with counterterm insertions (of course, these diagrams are at the same
order of the coupling constant as one--loop diagrams because of the counterterm
insertion). This remarkable claim is verified in \cite{CQT1}, where CQT explicitly
calculate the 10 counterterm diagrams and recover the
correct result for the four-point amplitude (see pp.~22-23 of \cite{CQT1})%
\footnote{In practice,
these authors choose to insert the self-energy result (\ref{selfenergy}) in the tree diagrams,
so what they compute is \emph{minus} the all--plus amplitude.}.

This result, apart from being very appealing in that one does not have to perform
any integrals (apart from the original integral that defined the counterterm)
so that the calculation reduces to tree--level combinatorics, will also turn out to be
a convenient starting point for performing the Mansfield transformation.
Specifically, our claim is that the whole series of all-plus amplitudes will
arise just from the counterterm action. In the following we will show how
this works explicitly for the four-point all-plus case, and then we will argue
for  the $n$-point case that the corresponding expression  derived from the counterterm 
has all the correct  singularities (soft and collinear), giving strong evidence that the result is true in general.

\section{The all-plus amplitudes from a counterterm} \label{allplus4section}

Having reviewed the relevant new features that arise when doing perturbation
theory with the worldsheet--motivated regulator of \cite{Thorn04}, we now have
all the necessary ingredients to perform the Mansfield change of variables on the
regulated lightcone Lagrangian. In this section, we will carry out this procedure.
We will first regulate lightcone self--dual Yang--Mills, which, as
discussed, will require us to introduce an explicit counterterm in the Lagrangian. Then
we will perform the Mansfield transformation on the original Lagrangian (converting
it to a free theory). We will  then  show that, upon inserting the change of variables
into the counterterm Lagrangian, we recover the all--plus amplitudes as vertices in the
theory.

\subsection{Mansfield transformation of $\Lcal_{\rm CT}$}

As we saw, the ``worldsheet-friendly'' regularisation requires us to add
a certain counterterm to the lightcone Yang--Mills action, required in order
to cancel the Lorentz-violating helicity--flipping gluon selfenergy. As mentioned
previously, the calculation of the all--plus amplitude 
can be tackled purely within the context of \emph{self-dual}
Yang--Mills, which  we will focus on from now on. We see that, as a result of
this regularisation, the complete action at the quantum level becomes:
\be
\Lcal_{\rm SDYM}^{(r)}=\Lcal_{+-}+\Lcal_{++-}+\Lcal_{\rm CT}
\ ,
\ee
where $\Lcal_{+-}+\Lcal_{++-}$ is the  classical  Lagrangian for 
self-dual Yang-Mills introduced in  \eqref{lclaga}. 
Although CQT do not
write down a spacetime Lagrangian for $\Lcal_{\rm CT}$, it is easy to see that
the following expression would have the right structure:
\be \label{LCTone}
\Lcal_{\rm CT}=-\frac{g^2N}{12\pi^2}\int_{\Sigma}\diff^{3} k^i\diff^{3} k^j \  A^i_{\;\;j}(k^i,k^j)
[(k_\zb^i)^2+(k_\zb^j)^2+k_\zb^i k_\zb^j]
A^j_{\;\;i}(k^j,k^i)
\ .
\ee
This expression depends explicitly on the dual, or region, momenta.
In (\ref{LCTone}) we have made use
of the simplest way to associate region momenta to fields,
which is to assign a region momentum to each \emph{index} line in
double--line notation \cite{tHooft74},
and thus a momentum $k^i$, $k^j$ to
each of the indices of the gauge field $A^i_{\;\;j}$ (now slightly extended into a dipole,
as would be natural from the worldsheet perspective, where an index is associated to each
boundary). Since each line has a natural orientation,
the actual momentum of each line can be taken to be the difference of the index momentum of
the incoming index line and the outgoing index line. So the momentum of $A^i_{\;\;j}(k^i,k^j)$ is
taken to be $p=k^j\!-\!k^i$.  As discussed above, this assignment can only be performed consistently
for planar diagrams, which is sufficient for our purposes.

Clearly, the structure of (\ref{LCTone}) is rather unusual. First of all, it depends
only on the antiholomorphic ($\zb$) components of the region momenta, and so is clearly
not (lightcone) covariant. Even more troubling is the fact that
it does not depend only on
\emph{differences} of region momenta, but also on their sums.
Since each field thus carries
more information than just its momentum, $\Lcal_{\rm CT}$ is a non--local object
from a four--dimensional point of view (although, as shown in \cite{CQT1}, it can be given a
perfectly local worldsheet description).

Leaving the above discussion as food for thought, we will now rewrite 
(\ref{LCTone}) in a more conventional way that is most convenient for
inserting into Feynman diagrams,  
\be \label{LCTtwo}
\Lcal_{\rm CT}=
-\frac{g^2N}{12\pi^2}\int_{\Sigma}\diff^{3} p\, \diff^{3} p' \, \delta(p+p')\
A^i_{\;\;j}(p')((k_\zb^i)^2+(k_\zb^j)^2+k_\zb^i k_\zb^j)
A^j_{\;\;i}(p)
\ .
\ee
In this expression, which can be thought of as the zero--mode or field theory limit 
of (\ref{LCTone}), all the region momentum dependence is confined to the polynomial
factor $(k_\zb^i)^2\!+\!(k_\zb^j)^2\!+\!k_\zb^i k_\zb^j$. This vertex, inserted into tree
diagrams, would exactly reproduce the effects of the counterterm pictured in (\ref{cteq}).
Although (\ref{LCTtwo}) still exhibits some of the apparently undesirable features
we discussed above, the calculations in \cite{CQT1} demonstrate that, after
summing over all possible insertions of this term, the final result is covariant and correctly
reproduces the all--plus amplitudes%
\footnote{Note that similar--looking treatments using index momenta
instead of line momenta for vertices, but which in the end sum up to covariant results have appeared
in the context of noncommutative geometry (see e.g.~\cite{MRS00}).
Although it is possible to write e.g. 
(\ref{LCTone}) in star--product form, at this stage it is not clear whether that is a useful
reformulation.}.
Therefore, we believe that its problematic properties
are really a virtue in disguise, and (as we will see explicitly)
they seem to be crucial in
obtaining the full series of $n$--point all--plus amplitudes from the
Mansfield transformation of a \emph{single} term.

We are now ready to perform the Mansfield change of variables.
In the spirit of the discussion earlier, we will
perform the transformation on the \emph{classical} part of the action only:
\be
\Lcal_{+-}(A,\bA)+\Lcal_{++-}(A,\bA)= \Lcal_{+-}(B,\bB)
\ee
Hence the classical part of the action has been converted to a free theory. Without
a regulator, this would be the whole story. However we now see that, within the
particular regularisation we are working with, the full Lagrangian $\Lcal_{\rm SDYM}^{(r)}$
contains one extra, one--loop piece, given by $\Lcal_{\rm CT}$ in (\ref{LCTtwo}), which is
quadratic in the positive helicity fields $A$. To complete the Mansfield
transformation, we will clearly need to expand this term in the new fields $B$,
using the Ettle--Morris coefficients (\ref{Aexpansion}).

Since $\Lcal_{\rm CT}$ depends only on the holomorphic $A$ fields, we will only
need the expansion of $A$ in terms of $B$ given in (\ref{Aexpansion}). As a first
check that $\Lcal_{\rm CT}$ leads to the right kind of structure, note that since
$A$ depends only on the holomorphic $B$ fields, all the new vertices are all--plus.
Thus, the full action, when expressed in terms of the $B$ fields, takes the schematic form:
\be
\Lcal_{\rm SDYM}^{(r)}(A,\bA)=\Lcal_{+-}(B,\bB)+\Lcal_{++}(B)+\Lcal_{+++}(B)+\Lcal_{++++}(B)+\cdots
\ee
In the next section we will calculate the four--point term $\Lcal_{++++}$ and demonstrate
that, when restricted on--shell, it reproduces the known form (\ref{allplus-4})
for the all--plus amplitude.

\subsection{The four--point case}

To begin with, we focus on the derivation of the four-point
all-plus vertex, whose on-shell version will give us
the four-point scattering amplitude. We will thus expand the old fields $A$ in the
counterterm \eqref{LCTtwo} (or \eqref{LCTone}) up to terms containing
four $B$-fields.

When inserting the Ettle--Morris coefficients into (\ref{LCTtwo}),
one has to sum over all possible cyclic orderings with which this can be done.
A complication  is that now the counterterm
itself depends on the ordering. In other words, we need to sum over all the ways
of assigning dual momenta to the indices. Schematically, the inequivalent terms
that we obtain are:
\be \label{AABBBB}
\begin{split}
AA&\ra (\sint B_1 B_2)(\sint B_3 B_4)+(\sint B_2 B_3)(\sint B_4 B_1)\\
&\quad+(\sint B_1 B_2 B_3) B_4
+(\sint B_2 B_3 B_4) B_1
+(\sint B_3 B_4 B_1) B_2
+(\sint B_4 B_1 B_2) B_3
\ ,
\end{split}
\ee
where the terms on the first line arise from doing two quadratic substitutions
and those on the second from doing one cubic substitution. All the other
possibilities are related by cyclicity of the trace. For definiteness, let us
now write down what one of these terms means explicitly:%
\footnote{We suppress the
overall factor of $-g^2N/(12\pi^2)$ until the end of this section. Also, the integrals
are implicitly taken to be on the quantisation surface $\Sigma$. }
\be \label{BBBBone}
\begin{split}
(\int B_1 &B_2 B_3) B_4\\
&=-2g^2\;\tr\int\diff p \diff p^4 \delta(p\!+\!p^4)\Big[\int\diff p^1\diff p^2\diff p^3
\delta(p\!-\!p^1\!-\!p^2\!-\!p^3)\frac{p_+}{\sqrt{p_+^1p_+^3}}\frac{1}{\ip{12}\ip{23}}\times\\
&\quad\times B(p^1)B(p^2)B(p^3)\Big]
\left[(k_\zb^3)^2+(k_\zb^4)^2+k_\zb^4 k_\zb^3\right]\; B(p^4)\\
&=2g^2\;\int\diff p^1\diff p^2\diff p^3\diff p^4\delta(p^1+p^2+p^3+p^4)\times\\
&\qquad\times\frac{p_+^4}{\sqrt{p_+^1p_+^3}}
\frac{(k_\zb^3)^2+(k_\zb^4)^2+k_\zb^4 k_\zb^3}
{\ip{12}\ip{23}}\tr\left[B(p^1)B(p^2)B(p^3)B(p^4)\right]
\ .
\end{split}
\ee
The reason this particular combination of $k_\zb$'s appears here is that, given
the ordering we chose, after the Mansfield transformation  the
counterterm ends up being on leg 4, and its line bounds
the regions with momenta $k_3$ and $k_3$. This is represented pictorially in Figure \ref{BBBBOne}.

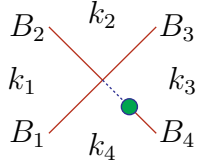
\begin{figure}[h]
\begin{center}
\begin{picture}(60,80)(0,0)
\put(0,0){
\SetColor{BrickRed}
\Line(10,10)(50,50)
\Line(10,50)(30,30)
\SetColor{Blue}
\DashLine(40,20)(30,30){1}
\SetColor{BrickRed}
\Line(40,20)(50,10)
\CCirc(40,20){3}{Blue}{Green}
\Text(0,30)[c]{$k_1$}
\Text(30,55)[c]{$k_2$}
\Text(60,30)[c]{$k_3$}
\Text(30,5)[c]{$k_4$}
\Text(2,10)[c]{$B_1$}
\Text(2,50)[c]{$B_2$}
\Text(58,10)[c]{$B_4$}
\Text(58,50)[c]{$B_3$}
}
\end{picture}
\caption{One of the contributions to the four--point all-plus vertex.}\label{BBBBOne}
\end{center}
\end{figure}

Although Figure \ref{BBBBOne} might suggest that there is a propagator
between the counterterm insertion and the location of the original $A$, which has now
split into three $B$'s, this is of course not the case since the whole expression
is a vertex at the same point. We have drawn the diagram in this fashion
to emphasise which leg the counterterm is located on after the
transformation. On the other hand, this vertex is nonlocal
(as discussed above, it was nonlocal even in the original variables, but
this is now compounded by the Mansfield coefficients, which contain momenta
in the denominator), so this notation serves as a useful reminder of that fact.

It is interesting to note that (\ref{BBBBone}) is essentially
the same expression as the sum of the two channels with the same
region momentum dependence that appear in CQT's calculation of this
amplitude using tree--level diagrammatics (compare with Eq.~83 in \cite{CQT1}), which
we illustrate in Fig. \ref{Tree}.
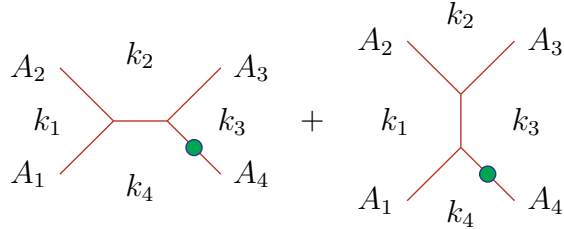
\begin{figure}[h]
\begin{center}
\begin{picture}(200,70)(0,0)
\put(0,0){
\SetColor{BrickRed}
\Line(10,10)(30,30)
\Line(10,50)(30,30)
\Line(30,30)(50,30)
\Line(50,30)(70,50)
\Line(50,30)(70,10)
\Text(0,30)[l]{$k_1$}
\Text(40,50)[b]{$k_2$}
\Text(80,30)[r]{$k_3$}
\Text(40,5)[u]{$k_4$}
\Text(5,10)[r]{$A_1$}
\Text(5,50)[r]{$A_2$}
\Text(75,10)[l]{$A_4$}
\Text(75,50)[l]{$A_3$}
\CCirc(60,20){3}{Blue}{Green}
}
\put(100,30){\Text(0,0)[l]{$+$}}
\put(130,-10){
\SetColor{BrickRed}
\Line(10,70)(30,50)
\Line(50,70)(30,50)
\Line(30,50)(30,30)
\Line(10,10)(30,30)
\Line(50,10)(30,30)
\Text(0,40)[l]{$k_1$}
\Text(30,75)[b]{$k_2$}
\Text(60,40)[r]{$k_3$}
\Text(30,5)[u]{$k_4$}
\Text(5,10)[r]{$A_1$}
\Text(5,70)[r]{$A_2$}
\Text(55,10)[l]{$A_4$}
\Text(55,70)[l]{$A_3$}
\CCirc(40,20){3}{Blue}{Green}
}
\end{picture}
\caption{The two diagrams with counterterm insertions on leg 4 that arise in the
calculation of CQT, and, combined, add up to the contribution in 
Fig. \ref{BBBBOne}.}\label{Tree}
\end{center}
\end{figure}
Thus we have a picture where one post--Mansfield transform vertex (with $B$'s)
effectively sums \emph{two} tree--level
pre--transformation (with $A$'s) Feynman diagrams. This is a first indication
that our calculation of the all--plus vertex can be mapped, practically
one--to--one, to that of the all--plus amplitude on pp.~22-23 of \cite{CQT1}.

Another type of contribution to the vertex arises when we transform \emph{both}
of the $A$'s in $\Lcal_{\rm CT}$. One of the two terms that we find is:
\be
\begin{split}
(\sint &B_2 B_3)(\sint B_4 B_1)\\
&=-2g^2\;\tr\int\diff p\, \diff p' \delta(p+p')\left[\int\diff p^2\diff p^3\delta(p-p^2-p^3)
\frac{p_+}{\sqrt{p_+^2p_+^3}}\frac{1}{\ip{23}}B(p^2)B(p^3)\right]\times\\
&\times \left((k_\zb^1)^2+(k_\zb^3)^2+k_\zb^1k_\zb^3\right)
\left[\int\diff p^4\diff p^1\delta(p'-p^4-p^1)\frac{p'_+}{\sqrt{p_+^4p_+^1}}
\frac{1}{\ip{41}}B(p^4)B(p^1)\right]\\
&=-2g^2\int\diff p^1\cdots \diff p^4\delta(p^1\!+\!p^2\!+\!p^3\!+\!p^4)
\frac{(p_+^2+p_+^3)(p_+^1+p_+^4)}{\sqrt{\pplfour}}
\frac{\left((k_\zb^1)^2+(k_\zb^3)^2+k_\zb^1k_\zb^3\right)}{\ip{23}\ip{41}}\\
&\qquad\qquad\times \tr\left[B(p^1)B(p^2)B(p^3)B(p^4)\right]
\ .
\end{split}
\ee
This contribution can also be mapped to one of the two terms with bubbles
on internal lines in CQT.

We can now tabulate all the terms that we obtain in this way by
making the schematic form (\ref{AABBBB}) precise.
Since the delta--function and trace over $B$ parts are the same for all these
terms, in Table \ref{Table} we just list the rest of the integrand.

\begin{table}[h]
\begin{center}
\begin{tabular}{|c|c|c|}\hline
Schematic form & Pictorial form & Integrand \\ \hline
$(\sint B_1B_2B_3)B_4$ &
\SetScale{0.5}
\begin{picture}(20,30)(0,0)
\put(-5,-5){
\SetColor{BrickRed}
\Line(10,10)(50,50)
\Line(10,50)(30,30)
\SetColor{Blue}
\DashLine(40,20)(30,30){1}
\SetColor{BrickRed}
\Line(40,20)(50,10)
\CCirc(40,20){3}{Blue}{Green}
}\end{picture}
&
$\frac{p_+^4}{\sqrt{p_+^1p_+^3}}\frac{k_3^2+k_4^2+k_3k_4}{\ip{12}\ip{23}}$\\ \hline
$(\sint B_2B_3B_4)B_1$ &
\SetScale{0.5}
\begin{picture}(20,30)(0,0)
\put(-5,-5){
\SetColor{BrickRed}
\Line(10,10)(20,20)
\Line(10,50)(50,10)
\SetColor{Blue}
\DashLine(20,20)(30,30){1}
\SetColor{BrickRed}
\Line(30,30)(50,50)
\CCirc(20,20){3}{Blue}{Green}
}\end{picture}
&
$\frac{p_+^1}{\sqrt{p_+^2p_+^4}}\frac{k_1^2+k_4^2+k_1k_4}{\ip{23}\ip{34}}$\\
\hline
$(\sint B_3B_4B_1)B_2$ &
\SetScale{0.5}
\begin{picture}(20,30)(0,0)
\put(-5,-5){
\SetColor{BrickRed}
\Line(10,10)(50,50)
\Line(10,50)(20,40)
\SetColor{Blue}
\DashLine(20,40)(30,30){1}
\SetColor{BrickRed}
\Line(30,30)(50,10)
\CCirc(20,40){3}{Blue}{Green}
}\end{picture}
&
$\frac{p_+^2}{\sqrt{p_+^3p_+^1}}\frac{k_1^2+k_2^2+k_2k_1}{\ip{34}\ip{41}}$\\
\hline
$(\sint B_4B_1B_2)B_3$ &
\SetScale{0.5}
\begin{picture}(20,30)(0,0)
\put(-5,-5){
\SetColor{BrickRed}
\Line(10,10)(30,30)
\Line(10,50)(50,10)
\SetColor{Blue}
\DashLine(40,40)(30,30){1}
\SetColor{BrickRed}
\Line(40,40)(50,50)
\CCirc(40,40){3}{Blue}{Green}
}\end{picture}
&
$\frac{p_+^3}{\sqrt{p_+^4p_+^2}}\frac{k_2^2+k_3^2+k_2k_3}{\ip{41}\ip{12}}$\\
\hline
$(\sint B_2 B_3)(\sint B_4 B_1)$ &
\SetScale{0.5}
\begin{picture}(20,30)(0,0)
\put(-5,-5){
\SetColor{BrickRed}
\Line(10,10)(30,20)
\Line(10,50)(30,40)
\SetColor{Blue}
\DashLine(30,20)(30,40){1}
\SetColor{BrickRed}
\CCirc(30,30){3}{Blue}{Green}
\Line(30,40)(50,50)
\Line(30,20)(50,10)
}
\end{picture}
& $-\frac{(p_+^2+p_+^3)(p_+^1+p_+^4)}{\sqrt{p_+^1p_+^2p_+^3p_+^4}}
\frac{k_1^2+k_3^2+k_1k_3}{\ip{23}\ip{41}}$
\\ \hline
$(\sint B_1 B_2)(\sint B_3 B_4)$ &
\SetScale{0.5}
\begin{picture}(20,30)(0,0)
\put(-5,-5){
\SetColor{BrickRed}
\Line(10,10)(20,30)
\Line(10,50)(20,30)
\SetColor{Blue}
\DashLine(20,30)(40,30){1}
\SetColor{BrickRed}
\CCirc(30,30){3}{Blue}{Green}
\Line(40,30)(50,50)
\Line(40,30)(50,10)
}
\end{picture}
&$-\frac{(p_+^3+p_+^3)(p_+^2+p_+^1)}{\sqrt{p_+^1p_+^2p_+^3p_+^4}}
\frac{k_4^2+k_2^2+k_4k_2}{\ip{34}\ip{12}}$
\\ \hline
\end{tabular}
\end{center}
\caption{The various contributions to the all--plus four--point vertex. Note that
we use the simplifying notation $k_i:=k_\zb^i$.} \label{Table}
\end{table}

To obtain the final form of the vertex, we are now instructed to sum over all
these contributions. Thus we can write
\be
\Lcal_{++++}(B)=2g^2\int\diff p^1\diff p^2\diff p^3 \diff p^4 \delta(p^1\!+\!p^2\!+\!p^3\!+\!p^4)
~\Vcal^{(4)}~\tr[B(p^1)B(p^2)B(p^3)B(p^4)]
\ee
where $\Vcal^{(4)}$ is given by the following expression:%
\footnote{For the sake of brevity  we omit a subscript $\bar{z}$ in the 
region momenta appearing in \eqref{initial}.}
\be\label{initial}
\begin{split}
\Vcal^{(4)}=&\frac{1}{\sqrt{\pplfour}}\frac{1}{\ip{12}\ip{23}\ip{34}\ip{41}}\times\\
&\times\bigg[p_+^4\sqrt{p_+^2p_+^4}(k_3^2+k_4^2+k_3k_4)\ip{34}\ip{41}
+p_+^1\sqrt{p_+^1p_+^3}(k_1^2+k_4^2+k_1k_4)\ip{12}\ip{41}\\
&+p_+^2\sqrt{p_+^2p_+^4}(k_2^2+k_1^2+k_2k_1)\ip{12}\ip{23}
+p_+^3\sqrt{p_+^3p_+^1}(k_3^2+k_2^2+k_2k_3)\ip{23}\ip{34}\\
&-(p_+^2+p_+^3)(p_+^1+p_+^4)(k_1^2+k_3^2+k_1k_3)\ip{12}\ip{34}\\
&-(p_+^3+p_+^4)(p_+^2+p_+^1)(k_4^2+k_2^2+k_4k_2)\ip{23}\ip{41}\bigg]
\; .
\end{split}
\ee
Comparing this to the expected answer (\ref{allplus-4}), we see that the
(quadratic) antiholomorphic momentum dependence should arise from the various
$k_\zb$ factors in (\ref{initial}). In \cite{CQT1}, CQT start from essentially
the same expression and demonstrate that it gives the correct result for the
all-plus amplitude. Therefore, following practically the same steps as those authors,
we can easily see that we obtain the expected answer. However, since we would
like to find the full vertex $\Vcal$, we will need to keep off--shell information, and
so we will choose a slightly different route.

The main complication in bringing (\ref{initial}) into a manageable form is
clearly the presence of the region momenta. We would like to disentangle their
effects as cleanly as possible. Therefore, our derivation will proceed by the
following steps:
\begin{enumerate}
\item First, we will show that (\ref{initial}) can be manipulated so that
the quadratic dependence on region momenta drops out, leaving only terms
linear in the region momenta.
\item Second, we will decompose the resulting expression into a part that
depends on the region momenta and one that does not. The $k$--dependent part
turns out to have a very simple form, and vanishes on--shell.
\item Finally, we will show that the $k$--independent part reduces to the
known amplitude.
\end{enumerate}

For the first step, we will need the following identity, which is proved in
appendix \ref{appdetails}:
\be \label{ident1}
\begin{split}
&p_+^4\sqrt{p_+^2p_+^4}\ip{34}\ip{41}
+p_+^1\sqrt{p_+^1p_+^3}\ip{12}\ip{41}
+p_+^2\sqrt{p_+^2p_+^4}\ip{12}\ip{23}
+p_+^3\sqrt{p_+^3p_+^1}\ip{23}\ip{34}\\
&-(p_+^2+p_+^3)(p_+^1+p_+^4)\ip{12}\ip{34}
-(p_+^3+p_+^4)(p_+^2+p_+^1)\ip{23}\ip{41}=0
\end{split}
\ee
Also, using the shorthand notation  $K_{ij}:=(k_\zb^i)^2+(k_\zb^j)^2+k_\zb^ik_\zb^j$:
we note the following very useful identity:
\be
K_{ij}=K_{ik}+(k_\zb^j-k_\zb^k)(k_\zb^i+k_\zb^j+k_\zb^k)=
K_{ik}+(k_\zb^j-k_\zb^k)l_{ijk}
\ee
where $1\leq k\leq n$ and $l_{ijk}=k_\zb^i+k_\zb^j+k_\zb^k$. Noting that, for
$j>k$, $k_\zb^j-k_\zb^k=p_\zb^{k+1}+p_\zb^{k+2}+\cdots p_\zb^j$, we can use this
to rewrite all the region momentum combinations appearing in (\ref{initial}) in
the following way:
\be\label{Kijdecomp}
\begin{split}
K_{34}&=\frac{1}{4}\left(K_{12}+K_{23}+K_{34}+K_{41}+(\bp_3+\bp_4)(l_{124}+l_{234})
+2(\bp_2+\bp_3)l_{134}\right)\\
K_{14}&=\frac{1}{4}\left(K_{12}+K_{23}+K_{34}+K_{41}-(\bp_2+\bp_3)(l_{134}+l_{123})
+2(\bp_3+\bp_4)l_{124}\right)\\
K_{12}&=\frac{1}{4}\left(K_{12}+K_{23}+K_{34}+K_{41}-(\bp_3+\bp_4)(l_{124}+l_{234})
-2(\bp_2+\bp_3)l_{123}\right)\\
K_{23}&=\frac{1}{4}\left(K_{12}+K_{23}+K_{34}+K_{41}+(\bp_2+\bp_3)(l_{134}+l_{123})
-2(\bp_3+\bp_4)l_{234}\right)\\
K_{13}&=\frac{1}{4}\left(K_{12}+K_{23}+K_{34}+K_{41}+(\bp_3-\bp_2)l_{123}
+(\bp_1-\bp_4)l_{134}\right)\\
K_{24}&=\frac{1}{4}\left(K_{12}+K_{23}+K_{34}+K_{41}+(\bp_4-\bp_3)l_{234}
+(\bp_2-\bp_1)l_{124}\right)\\
\end{split}
\ee
where we have introduced the notation $\bp_i=p_\zb^i$. We have thus expressed
all the quadratic region momentum dependence in terms of the common factor
$K_{12}+K_{23}+K_{34}+K_{41}$, and, given (\ref{ident1}), it is clear that this
contribution will vanish.%
\footnote{One
could have chosen a different combination of the $K_{ij}$'s, but we find the
symmetric choice in (\ref{Kijdecomp}) convenient.}

After this step, we are left with an expression which is linear in the
region momenta. We will now proceed in a similar way, and rewrite all the
expressions that contain $l_{ijk}$ in terms of a suitably chosen common factor:
\be \label{lijkdecomp}
\begin{split}
&l_{124}+l_{234}=\frac{3}{2}(k_\zb^1+k_\zb^2+k_\zb^3+k_\zb^4)-\half(p_\zb^1+p_\zb^3)\\
&l_{134}+l_{123}=\frac{3}{2}(k_\zb^1+k_\zb^2+k_\zb^3+k_\zb^4)-\half(p_\zb^2+p_\zb^4)\\
&2l_{234}=\frac{3}{2}(k_\zb^1+k_\zb^2+k_\zb^3+k_\zb^4)+\half(2p_\zb^2+p_\zb^3-p_\zb^1)\\
&2l_{123}=\frac{3}{2}(k_\zb^1+k_\zb^2+k_\zb^3+k_\zb^4)+\half(2p_\zb^1+p_\zb^2-p_\zb^4)\\
&2l_{134}=\frac{3}{2}(k_\zb^1+k_\zb^2+k_\zb^3+k_\zb^4)+\half(2p_\zb^3+p_\zb^4-p_\zb^2)\\
&2l_{124}=\frac{3}{2}(k_\zb^1+k_\zb^2+k_\zb^3+k_\zb^4)+\half(2p_\zb^4+p_\zb^1-p_\zb^3)\\
\end{split}
\ee
In appendix \ref{appdetails} we show that the total coefficient of the
common $(k_\zb^1+k_\zb^2+k_\zb^3+k_\zb^4)$ factor is
\be \label{ident2}
\begin{split}
&\frac{3}{8}\left[p_+^4\sqrt{p_+^2p_+^4}(+(\bp_3+\bp_4)
+(\bp_2+\bp_3))\ip{34}\ip{41}
+p_+^1\sqrt{p_+^1p_+^3}(-(\bp_2+\bp_3)
+(\bp_3+\bp_4))\ip{12}\ip{41}\right.\\
&+p_+^2\sqrt{p_+^2p_+^4}(-(\bp_3+\bp_4)
-(\bp_2+\bp_3))\ip{12}\ip{23}
+p_+^3\sqrt{p_+^3p_+^1}(+(\bp_2+\bp_3)
-(\bp_3+\bp_4))\ip{23}\ip{34}\\
&-(p_+^2+p_+^3)(p_+^1+p_+^4)(\half(\bp_3-\bp_2)+\half(\bp_1-\bp_4))\ip{12}\ip{34}\\
&-(p_+^3+p_+^4)(p_+^2+p_+^1)(\half(\bp_4-\bp_3)+\half(\bp_2-\bp_1))\ip{23}\ip{41}
\bigg]=\\
&=-\frac{3}{16}[(12)+(23)+(34)+(41)]\sum_{i=i}^4\frac{(p_i)^2}{p_+^i}
\ , 
\end{split}
\ee
where $(p_i)^2$ is the full covariant momentum squared, and $(ij)=p_+^ip_z^j-p_+^jp_z^i$.
Thus we see that the complete dependence on the region momenta can be
rewritten as follows:
\be \label{regionvertex}
\Vcal^{(4)}_{k}=-\frac{3}{16}\frac{(12)+(23)+(34)+(41)}{\ip{12}\ip{23}\ip{34}\ip{41}}~\left[\sum_{i=1}^4 ~k_\zb^i\right]~\sum_{i=i}^4\frac{(p_i)^2}{p_+^i}
\ .
\ee
It is rather satisfying that the region momentum dependence of the vertex takes
this simple form, which clearly vanishes when the external legs are on--shell, and
thus will not contribute to the all--plus amplitudes.

Having completely disentangled the region momenta $k_\zb$ from the actual momenta
$p_\zb$, we will now focus on the terms containing only the latter, which were produced
during the decompositions in (\ref{lijkdecomp}). After a few simple manipulations,
they can be rewritten as\footnote{We write $V^{(4)}=\sqrt{\pplfour}\ip{12}\ip{23}\ip{34}\ip{41}\Vcal^{(4)}$.} 
\be \label{peqsimp}
\begin{split}
V^{(4)}_p=\frac{1}{8}&\bigg[
p_+^4\sqrt{p_+^2p_+^4}[(\bp_1+\bp_2)(\bp_1-\bp_2)+(\bp_3+\bp_2)(\bp_3-\bp_2)]\ip{34}\ip{41}\\
&+p_+^1\sqrt{p_+^1p_+^3}[(\bp_2+\bp_3)(\bp_2-\bp_3)+(\bp_4+\bp_3)(\bp_4-\bp_3)]\ip{41}\ip{12}\\
&+p_+^2\sqrt{p_+^2p_+^4}[(\bp_3+\bp_4)(\bp_3-\bp_4)+(\bp_1+\bp_4)(\bp_1-\bp_4)]\ip{12}\ip{23}\\
&+p_+^3\sqrt{p_+^3p_+^1}[(\bp_4+\bp_1)(\bp_4-\bp_1)+(\bp_2+\bp_1)(\bp_2-\bp_1)]\ip{23}\ip{34}\\
&-(p_+^2+p_+^3)(p_+^1+p_+^4)[(\bp_3-\bp_2)(\bp_1-\bp_4)-(\bp_1+\bp_2)^2]\ip{12}\ip{34}\\
&-(p_+^3+p_+^4)(p_+^2+p_+^1)[(\bp_4-\bp_3)(\bp_2-\bp_1)-(\bp_2+\bp_3)^2]\ip{23}\ip{41}
\bigg]
\ .
\end{split}
\ee
This expression, together with (\ref{regionvertex}) is our proposal for the off--shell
four--point all--plus vertex that should be part of the MHV-rules formalism at the
quantum level. It would be very interesting to elucidate its structure
and bring it into a more compact form. For the moment, however, we will be content to
demonstrate that (\ref{peqsimp}) is equal on shell  to the sought--for amplitude. 

To that end, we will follow a similar approach to CQT, and rewrite all the holomorphic spinor
brackets in terms of the following three: $\ip{12}\ip{34},\ip{23}\ip{41},\ip{12}\ip{41}$.
To achieve this, we use momentum conservation and a certain cyclic identity
(see appendix \ref{notation}) to write
\be
\begin{split}
p_+^4\sqrt{p_+^2p_+^4}\ip{34}\ip{41}&=
p_+^4\sqrt{p_+^4}\left(-\sqrt{p_+^3}\ip{42}-\sqrt{p_+^4}\ip{23}\right)\ip{41}\\
&=\left[-p_+^4\sqrt{p_+^3p_+^4}\ip{42}-(p_+^4)^2\right]\ip{41}\\
&=\left[-p_+^4\sqrt{p_+^3}\left(-\sqrt{p_+^1}\ip{12}-\sqrt{p_+^3}\ip{32}\right)
-(p_+^4)^2\ip{23}\right]\ip{41}\\
&=p_+^4\sqrt{p_+^3p_+^1}\ip{12}\ip{41}-p_+^4(p_+^4+p_+^3)\ip{23}\ip{41}
\ .
\end{split}
\ee
In a similar way, we can show that
\be
\begin{split}
&p_+^2\sqrt{p_+^2p_+^4}\ip{12}\ip{23}=
p_+^2\sqrt{p_+^3p_+^1}\ip{12}\ip{41}
-p_+^2(p_+^2+p_+^3)\ip{34}\ip{12}\;,\\
&p_+^3\sqrt{p_+^1p_+^3}\ip{23}\ip{34}=
\!-\!\left[p_+^3(p_+^3\!+\!p_+^2)\ip{12}\ip{34}-p_+^3(p_+^1\!+\!p_+^2)\ip{23}\ip{41}
\!+\!p_+^3\sqrt{p_+^1p_+^3}\ip{12}\ip{14}\right]\;.
\end{split}
\ee
Collecting all the terms together, and manipulating the resulting expressions,
it is straightforward to show that (\ref{peqsimp}) simplifies to just
\be\label{intermediate1}
\begin{split}
V^{(4)}_p=\frac{1}{4}\bigg[&\ip{23}\ip{41}\{34\}(p_+^1+p_+^2)[(\bp_1-\bp_2)-(\bp_2+\bp_3)]\\
&\!+\!\ip{12}\ip{34}\{23\}(p_+^2+p_+^3)[(\bp_1+\bp_2)+(\bp_1-\bp_4)]\\
&\!+\!\ip{12}\ip{41}\sqrt{p_+^3p_+^1}\big[(\bp_1+\bp_2)(\{41\}+\{32\})
\!+\!(\bp_2+\bp_3)(\{12\}+\{43\})\big]\bigg]\;,
\end{split}
\ee
where we use the notation \cite{EttleMorris}
$\{ij\}=p_+^ip_\zb^j-p_+^jp_\zb^i=(1/\sqrt{2}) \sqrt{p_+^ip_+^j}[ij]$.
Converting
to the usual antiholomorphic bracket notation, we rewrite (\ref{intermediate1}) as
\be \label{intermediate}
\begin{split}
V^{(4)}_p=\frac{1}{4\sqrt{2}}\bigg[&\ip{23}\ip{41}\sqrt{p_+^3p_+^4}[34](p_+^1+p_+^2)
[(\bp_1\!-\!\bp_2)-(\bp_2\!+\!\bp_3)]\\
&+\ip{12}\ip{34}\sqrt{p_+^2p_+^3}[23](p_+^2+p_+^3)[(\bp_1\!+\!\bp_2)+(\bp_1\!-\!\bp_4)]\\
&+\ip{12}\ip{41}\big[(\bp_1+\bp_2)(p_+^1\sqrt{p_+^3p_+^4}[41]+p_+^2\sqrt{p_+^2p_+^1}[32])\\
&+(\bp_2+\bp_3)(p_+^1\sqrt{p_+^2p_+^3}[12]+p_+^3\sqrt{p_+^1p_+^4}[43])\big]\bigg]
\ .
\end{split}
\ee
Note that so far this expression is completely off shell. We will now
show that on shell it reduces to the known result (\ref{allplus-4}). In
doing this we will  keep track of the $p^2$ terms that appear
when applying momentum conservation in the form
\be
\sum_k \ip{ik}[kj]= \sqrt{p_+^ip_+^j}\sum_k \frac{(p_k)^2}{p_+^k}
\ .
\ee
These terms are collected in appendix \ref{appdetails}. 

We start by  rewriting each of the terms in the last two lines of (\ref{intermediate}) 
as follows
\be \label{decomp}
\begin{split}
&\ip{12}\ip{41}[41]~p_+^1\sqrt{p_+^3p_+^4}(\bp_1+\bp_2)=
-\ip{23}\ip{41}[34]~p_+^1\sqrt{p_+^3p_+^4}(\bp_1+\bp_2)\\
&\ip{12}\ip{41}[32]~p_+^3\sqrt{p_+^1p_+^2}(\bp_1+\bp_2)
=-\ip{12}[32]\ip{42}p_+^2p_+^3(\bp_1+\bp_2)\\
&\phantom{\ip{12}\ip{41}[32]~p_+^3\sqrt{p_+^1p_+^2}(\bp_1+\bp_2)=}
-\ip{12}\ip{34}[23]~p_+^3\sqrt{p_+^2p_+^3}(\bp_1+\bp_2)\\
&\ip{12}\ip{41}[12]~p_+^1\sqrt{p_+^2p_+^3}(\bp_2+\bp_3)=
-\ip{12}\ip{34}[23]~p_+^1\sqrt{p_+^2p_+^3}(\bp_2+\bp_3)\\
&\ip{12}\ip{41}[43]~p_+^3\sqrt{p_+^1p_+^4}(\bp_2+\bp_3)
=-\ip{41}\ip{23}[34]~p_+^3\sqrt{p_+^3p_+^4}(\bp_2+\bp_3)\\
&\phantom{\ip{12}\ip{41}[43]~p_+^3\sqrt{p_+^1p_+^4}(\bp_2+\bp_3)=}
-\ip{41}[43]\ip{42}p_+^4p_+^3(\bp_2+\bp_3)
\, . 
\end{split}
\ee
We also transform the $\ip{12}\ip{34}$ term using the Schouten identity and also
momentum conservation, 
\be \label{prefinal}
\ip{12}\ip{34}[23]\sqrt{p_+^2p_+^3}\!=\!
\ip{23}\ip{41}[34]\sqrt{p_+^3p_+^4}\!+\!\ip{14}\ip{23}[13]\sqrt{p_+^1p_+^3}
\!-\!\ip{13}\ip{42}[23]\sqrt{p_+^2p_+^3}
\ , 
\ee
and add up all contributions to the $\ip{23}\ip{41}$ term, which are
\be \label{final}
\begin{split}
\frac{1}{4\sqrt{2}}\ip{23}\ip{41}[34]\sqrt{p_+^3p_+^4}&
\big[4(p_+^2\bp_1-p_+^1\bp_2)+2(p_+^3\bp_1-p_+^1\bp_3)\big]\\
=\frac{1}{4\sqrt{2}}\ip{23}\ip{41}[34]\sqrt{p_+^3p_+^4}&[4\{21\}+2\{31\}]\;.
\end{split}
\ee
Converting to the spinor bracket, the first of these terms is 
\be \label{finalfinal}
-\half\sqrt{\pplfour}[12]\ip{23}[34]\ip{41}
\ , 
\ee
while the remaining terms from (\ref{decomp}) and (\ref{prefinal}) combine to give
\be \label{otherterms}
\begin{split}
&\left(\ip{14}\ip{23}[13]\sqrt{p_+^1p_+^3}-\ip{13}\ip{42}[23]\sqrt{p_+^2p_+^3}\right)
(p_+^2+p_+^3)[(\bp_1\!+\!\bp_2)+(\bp_1\!-\!\bp_4)]\\
&+\ip{12}[32]\ip{42}p_+^2[p_+^2(\bp_1+\bp_2)-p_+^4(\bp_2+\bp_3)]\\
=&-\ip{14}[13]\ip{12}p_+^3(p_+^2+p_+^3)[(\bp_1\!+\!\bp_2)+(\bp_1\!-\!\bp_4)]\\
&+\ip{12}[32]\ip{42}p_+^2[p_+^2(\bp_1+\bp_2)-p_+^4(\bp_2+\bp_3)]\\
=&-\ip{14}[13]\ip{12}p_+^3(2(p_+^2+p_+^3)\bp_1-2p_+^1(\bp_2+\bp_3))=2\ip{14}[13]\ip{12}p_+^3\{41\} 
\end{split}
\ee
(where we suppress an overall $1/(4\sqrt2)$)
and we see that (\ref{otherterms}) cancels the second term in (\ref{final}), thus
showing that (\ref{finalfinal}) is the complete on-shell answer. Reintroducing all 
the prefactors, we thus find that the amplitude is
\be
\begin{split}
\Acal^{(4)}&=-\frac{g^2N}{12\pi^2}
\frac{2g^2}{\sqrt{\pplfour}}\frac{1}{\ip{12}\ip{23}\ip{34}\ip{41}}\times
\left[-\half\sqrt{\pplfour}[12]\ip{23}[34]\ip{41}\right]\\
&=\frac{g^4N}{12\pi^2}\frac{[12][34]}{\ip{12}\ip{34}}
\ .
\end{split}
\ee
Now note that, as discussed in appendix \ref{notation}, in order to convert to the usual 
Yang--Mills theory normalisation we need to send $g\ra g/\sqrt{2}$. We
conclude that $\Acal^{(4)}$ gives precisely the result (\ref{allplus-4}) for the all--plus
scattering amplitude.

\subsection{The general all--plus amplitude} \label{Section:Npoint}

We have just given an explicit derivation of the four point
all-plus amplitude, from the two-point counterterm (\ref{LCTtwo}).
We will argue in the following that  this two-point counterterm
contains {\it all} the all-plus amplitudes.

First, we can see immediately that the counterterm
(\ref{LCTtwo}) has the right kind of structure.
Consider the $n$--point all--plus amplitude \cite{BCDK93}:
\be \label{allplus-N}
{\cal{A}}^{(n)}=\sum_{1\leq i<j<k<l\leq n}\frac{\ip{ij}[jk]\ip{kl}[li]}{\ip{12}\cdots \ip{n1}}
\ . 
\ee
In terms of spinor brackets this amplitude has terms of the form $\ip{\ \ }^{2-n} [\ \ ]^2$.
A quick look at the Ettle-Morris coefficients shows that, for an $n$--point
vertex coming from $\Lcal_{\rm CT}$, they contribute exactly $2-n$ powers of
the spinor brackets $\ip{\ \ }$. Furthermore, there are
exactly two powers of $[\ \ ]$ coming from the counterterm Lagrangian
$\Lcal_{\rm CT}\sim(k_\zb^2)A^2$ -- one for each power of $k$. Thus the
general structure of $\Lcal_{\rm CT}$ is appropriate to reproduce (\ref{allplus-N}).

Pictorially, we can represent the general $n$--point amplitude, arising from
the counterterm in the new variables, as in Figure \ref{NPointfigure}.

\begin{figure}[h]
\begin{center}
\begin{picture}(80,110)(0,0)
\put(0,0){
\SetColor{BrickRed}
\Line(10,20)(30,40)
\Line(20,15)(30,40)
\Line(50,20)(30,40)
\Line(30,70)(50,90)
\Line(30,70)(40,95)
\Line(30,70)(10,90)
\SetColor{Blue}
\DashLine(30,40)(30,70){1}
\CCirc(30,55){3}{Blue}{Green}
\SetColor{Green}
\DashCArc(30,40)(25,-95,-55){1}
\DashCArc(30,40)(24,-95,-55){1}
\DashCArc(30,70)(25,85,125){1}
\DashCArc(30,70)(24,85,125){1}
\Text(5,24)[tr]{$B_i$}
\Text(20,10)[t]{$B_{i-1}$}
\Text(53,24)[tl]{$B_{j+1}$}
\Text(60,86)[b]{$B_{j}$}
\Text(40,100)[b]{$B_{j-1}$}
\Text(5,95)[r]{$B_{i+1}$}
\Text(10,55)[r]{$k_{i}$}
\Text(50,55)[l]{$k_{j}$}
}
\end{picture}
\caption{The structure of a generic term contributing to the $n$--point vertex.
All momenta are taken to be outgoing, and all indices are modulo $n$.}\label{NPointfigure}
\end{center}
\end{figure}
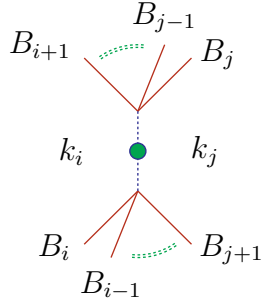

Thus we can write this $n$--point all--plus vertex as follows:
\be \label{npoint}
\begin{split}
\Acal^{(n)}_{+\cdots +}=&\int_{1\cdots n}\delta(p+p')
\sum_{1\leq i<j\leq n}\Yrm(p;j+1,\ldots,i)
\left((k_\zb^{i})^2+(k_\zb^{j})^2+k_\zb^{i}k_\zb^{j}\right)
\Yrm(p';i+1,\ldots, j)\times\\ \\
&\times\tr[B_iB_{i+1}\cdots B_jB_{j+1}\cdots B_{i-1}]\\ \\
=& (\sqrt2i)^{n-2}\int_{1\cdots n}\!\delta(p^1\!+\!\cdots\! +\!p^n)
\sum_{1\leq i<j\leq n}\frac{(p_+^{j+1}+\cdots+ p_+^{i})}{\sqrt{p_+^{j+1} p_+^{i}}}
\frac{1}{\ip{j+1,j+2}\cdots\ip{i-1,i}}\times\\ \\
&\times\left((k_\zb^{i})^2+(k_\zb^{j})^2+k_\zb^{i}k_\zb^{j}\right)
\frac{(p_+^{i+1}+\cdots+ p_+^{j})}{\sqrt{p_+^{i+1} p_+^{j}}}
\frac{1}{\ip{i+1,i+2}\cdots \ip{j-1,j}}\tr[B_1\cdots B_n]
\ .
\end{split}
\ee
Focusing only on the relevant part of the above expression, and ignoring all
coefficients, the general structure we obtain is the following:
\be \label{allplusproposed}
\Vcal^{(n)}_{+\cdots+}=\frac{1}{\ip{12}\cdots \ip{n1}}\!\times\!\!\left[\sum_{1\leq i<j\leq n}
\frac{\ip{j,j+1}\ip{i,i+1}}{\sqrt{p_+^ip_+^{i+1}p_+^jp_+^{j+1}}}
(k_+^{j}-k_+^{i})^2((k_\zb^{i})^2
+(k_\zb^{j})^2+k_\zb^{i}k_\zb^{j})\right]
\ee
where we have extracted the denominator at the expense of introducing the two missing
holomorphic factors $\ip{j,j+1}$ and $\ip{i,i+1}$ in the numerator. We also made use
of the fact that
\be
k^{j}-k^{i}=p^{i+1}+p^{i+2}+\cdots+ p^{j}=-(p^{j+1}+p^{j+2}+\cdots+ p^{i})\;,
\ee
applied to the $+$ components, to rewrite the two $p_+$ sums in the numerator
in terms of the $k$'s (this gives rise to a minus which we suppress).

It is easy to verify that, for $n=4$, this sum reproduces the 6 contributions
that appeared in the four--point case, and (as we explicitly showed above) combined
to give the expected answer. Therefore, we would like to propose that the vertex
(\ref{allplusproposed}) will reduce on--shell to an expression proportional
to (\ref{allplus-N}). We will not attempt to prove this statement
here%
\footnote{It is perhaps interesting to remark that the proof would
involve converting the double sum in (\ref{allplusproposed}) to the quadruple
sum in (\ref{allplus-N})---a state of affairs which has appeared before in a rather
different context \cite{Bern:1996ja}.}, but will
instead move on to study the general properties of the $n$-point expression
(\ref{npoint}).

Whilst the explicit calculation for the four point case was rather involved
as we saw earlier, the study of the general
properties of the $n$--point amplitudes proves much simpler. In particular, we
will show that the collinear and soft limits of the expressions proposed for
the $n$--point case can be very easily shown to be correct.
Let us start by introducing some simplifying notation. One can write the
change of variables for the $A$ field as
\be\label{MandM}
A_1 = \Yrm_{12}B_2 + \Yrm_{123}B_2B_3 + \Yrm_{1234}B_2B_3B_4 + \cdots,
\ee
where
\be \label{MM1}
\Yrm_{12} = \delta_{12}, \qquad \Yrm_{123} = \frac{1_+}{(23)}, \qquad
             \Yrm_{1234} = \frac{1_+3_+}{(23)(34)},
\ee
and generally
\be\label{Ys}
\Yrm_{12\dots n} = \frac{1_+3_+4_+\dots(n-1)_+}{(23)(34)\dots(n-1\ n)}
\
\ee
(for simplicity, we are dropping inconsequential constant
factors in this discussion).
This notation is similar to that of \cite{EttleMorris}. 
Integrations and the insertion of suitable delta functions are understood, and
can be illustrated by comparing the short-hand expressions above with the
full equations given earlier.
It will prove convenient to define
\be\label{Ks}
K_{ij} = k_i^2 + k_j^2 + k_i k_j, \qquad k_i:= k_\zb^i.
\ee
We will use the expression $\Yrm_{\bullet 12\dots n}$ in the following, where the
dot in the first placemark in the $\Yrm$ means that one substitutes in that place
 the negative of the sum of the other momenta.
Then the result which we have proved above for the four point amplitude
$V_{1234}$ can be expressed as
\be\label{4pointA}
\begin{split}
V_{1234} =& K_{43}\Yrm_{\bullet 4} \Yrm_{\bullet 123} 
+ K_{14} \Yrm_{\bullet 1} \Yrm_{\bullet 234}
+ K_{21}\Yrm_{\bullet 2} \Yrm_{\bullet 341} 
+ K_{32} \Yrm_{\bullet 3} \Yrm_{\bullet 412}\\
&+ K_{31}\Yrm_{\bullet 23} \Yrm_{\bullet 41} 
+ K_{24} \Yrm_{\bullet 12} \Yrm_{\bullet 34}\ ,
\end{split}
\ee
or very simply
\be\label{4point}
V_{1234} = \sum_{1\leq i<j\leq 4} K_{i j}\Yrm_{\bullet\, j+1 \dots i} \Yrm_{\bullet\, i+1\dots j}\ .
\ee

It is clear that the general conjecture that all the $n$--point all plus amplitudes
are generated from the two-point counterterm \eqref{LCTtwo} translates into
the proposal that the $n$-point all-plus amplitude $V_{12\dots n}$ is given by
\be\label{conj}
V_{12\dots n} = \sum_{1\leq i<j\leq n} 
K_{i j}\Yrm_{\bullet\, j+1 \dots i} \Yrm_{\bullet\, i+1\dots j}\ ,
\ee
Let us now show that the expression on the right-hand side of \eqref{conj}
has precisely the same soft and collinear limits as the known amplitude on the
left-hand side.

\noindent{\bf Collinear limits}

Under the collinear limit
\be\label{collinear1}
p_i \rightarrow z P\ , \qquad p_{i+1} \rightarrow (1-z)P\ , 
\qquad P^2 \to 0 \ , 
\ee
the $n$-point amplitude $V_{12\dots n}$ behaves as
\be\label{collinear2}
V_{12\dots n} \rightarrow  \frac{1}{z(1-z)} \frac{i_+}{(i\, i+1)}\
                 V_{12\dots i\ i+2\dots n} \ ,
\ee
where we relabel $P\rightarrow p_i$ after the limit is taken
(the $i_+$ and $(i\ i+1)$ factors involve momenta rather than spinors, which
is why the $z$-dependent factor is $1/z(1-z)$, rather than the conventional
$1/\sqrt{z(1-z)}$).

Consider the behaviour of the right-hand side of \eqref{conj} under the limit
\eqref{collinear1}.
The first point is that if the indices $i, i+1$ lie on different $\Yrm$'s, then
there are no poles generated in this collinear limit.  This is clear from the
explicit expressions for the $\Yrm$'s in \eqref{Ys}. Thus we may ignore any terms of this type.
It is then immediate from the explicit forms of the $\Yrm$'s that
\be\label{Ycollinear}
\Yrm_{12\dots s} \rightarrow  \frac{1}{z(1-z)} \frac{i_+}{(i\, i+1)}\
                 \Yrm_{12\dots i\ i+2\dots s} \ ,
\ee
for any $i = 2,\dots s-1$, with $s\leq n$ (the first index in $\Yrm$
never contributes in a collinear
limit, as one can see from the conjecture \eqref{conj}). Thus we see that the $\Yrm$ expressions
have the right sort of collinear behaviour. It is straightforward to see that the
$K$ coefficients in \eqref{conj} also get relabelled correctly  in the collinear limit;
they are not explicitly involved as they refer to pairs of momenta attached to different
$\Yrm$ fields, and as we saw, these do not contribute.

It is then immediate to see that the summation over the products of $\Yrm$'s in
\eqref{conj} reduces correctly in the collinear limit to the required summation over
products of $\Yrm$'s with one fewer leg in total.
Hence the proposal \eqref{conj}
 for the amplitude has precisely the same collinear limits
as the physical amplitude.

\noindent{\bf Soft limits}

We also find that there is a simple derivation of the soft limits of the expression in
\eqref{conj}. In the soft limit
\be\label{soft1}
p_j \rightarrow 0 
\ ,
\ee
the $n$-point amplitude $V_{12\dots n}$ behaves as
\be\label{soft2}
V_{12\dots n} \rightarrow S(j)\
                 V_{12\dots j-1\ j+1\dots n} \ ,
\ee
where we assume cyclic ordering as usual, so that, for example, $p_{n+1}=p_1$.
The soft function $S(j)$ is given in terms of the momentum brackets by
\be\label{softS}
S(j) = \frac{j_+ (j-1\, j+1)}{(j-1 \,j)\,(j\, j+1)}\ .
\ee
The $\Yrm$ functions have a simple behaviour under soft limits. One has immediately
that in the soft limit $p_j\rightarrow 0$,
\be\label{Ysoft}
\Yrm_{12\dots s} \rightarrow  S(j)\
                 \Yrm_{12\dots j-1\ j+1\dots s} \ ,
\ee
for $j=3,\dots s-1$ (with $s\leq n$). For the soft limits corresponding to
the case missing in the above, we need the results
\be\label{interm1}
\Yrm_{\bullet s+1\dots j} =  \Yrm_{\bullet s+1\dots j-1}\, \frac{(j-1)_+}{(j-1\, j)},
\qquad
\Yrm_{\bullet j\dots s} =  \Yrm_{\bullet j+1\dots s}\, \frac{(j+1)_+}{(j\, j+1)} \ ,
\ee
which follow from the definitions of the $\Yrm$'s, and
\be\label{interm2}
\frac{(j+1)_+}{(j\, j+1)} + \frac{(j-1)_+}{(j-1\, j)} =
 \frac{j_+(j-1\, j+1)}{(j-1\, j)\ (j\, j+1)} = S(j)
\ ,
\ee
which follows from the cyclic identity $i_+(jk) + j_+(ki) + k_+(ij) = 0$.
Finally, from relabelling the $K$'s we have in the soft limit that
$K_{sj}\rightarrow K_{s j-1}$.
Then it follows that in the soft limit
\be\label{lala}
K_{sj}\ \Yrm_{\bullet s+1\dots j}\ \Yrm_{\bullet j+1\dots s} +
K_{sj-1}\ \Yrm_{\bullet s+1\dots j-1}\ \Yrm_{\bullet j\dots s}
\rightarrow
S(j)
K_{s j-1}\ \Yrm_{\bullet s+1\dots j-1}\ \Yrm_{\bullet j+1\dots s} \ ,
\ee
as required.

Again, it is then easy to see that the summation over the products of $\Yrm$'s in
\eqref{conj} reduces correctly in the soft limit to the required summation over
products of $\Yrm$'s with one fewer leg in total.
Hence the proposal \eqref{conj} for the amplitude has precisely the same soft limits
as the physical amplitude.

\section{Discussion}

Whilst new, twistor-inspired methods for calculating amplitudes in gauge theory have
led to much progress, the lack of a systematic action-based formulation
which incorporates these new ideas has been an impediment to further developments.
MHV diagrams have the two advantages of being
closely allied to the twistor picture, as well as providing an explicit realisation of the
dispersion and phase space integrals fundamental to unitarity-based methods. However,
without an action formalism, standard MHV methods have so far been mainly restricted to
massless theories at one-loop level, and to the cut-constructible parts of amplitudes.

The advent of a classical MHV Lagrangian for gauge theory, derived from lightcone
YM theory \cite{Gorsky:2005sf,Mansfield,EttleMorris},
provides the basis for transcending these limitations. In order for this
to be realised, it is necessary to describe the quantum MHV theory.
What we have done in this paper is to investigate this quantum theory.
Using the regularisation methods of \cite{Thorn05,CQT1,CQT2}, we
have provided arguments that the simplest one-loop counterterm in the
quantum MHV theory -- a two point vertex -- provides an extraordinarily
concise generating function for the infinite sequence of one-loop, all-plus
helicity amplitudes in YM theory.
We showed this by explicit calculation for the four-point case,
and then proved that the soft and collinear limits of the conjectured
$n$-point amplitude precisely matched those of the correct answer.

We would like to emphasise that the simplicity of our approach --- which reduced
the calculations of the loop amplitudes we considered to tree--level algebraic 
manipulations--- is largely due to the four--dimensional nature of the 
regularisation scheme we employed. By staying in four
dimensions, we preserve the appealing features of the inherently  
four--dimensional field redefinition of \cite{Gorsky:2005sf,Mansfield}.

Based upon this result, it is very natural to conjecture that the full quantum
YM theory is correctly described by this quantum MHV Lagrangian. The
correct ingredients appear to be present. For example, in the approach of 
\cite{Thorn05,CQT1,CQT2} there arise one-loop counterterms with
helicities $(++), (++-), (--), (--+)$. We studied the $(++)$ counterterm in
this paper, arguing that when expressed in the $(B, \Bbar)$ variables this
generates the full set of all-plus amplitudes. Transforming the $(++-)$ counterterm
to $(B, \Bbar)$ variables will generate an infinite sequence of single--minus vertices.
There will be other contributions to single-minus vertices from combinations of
all-plus vertices and MHV vertices.
It would be surprising if the combined contributions of these did not
lead to the correct YM single-minus expressions. Certainly all of these
have the correct powers of spinor brackets for this to be the case.

Transforming the $(--)$ and $(--+)$ counterterms to $(B, \Bbar)$ variables will lead to new
contributions to MHV vertices\footnote{In the MHV case there are additional counterterms 
noted in \cite{CQT2} which may also need to be taken into account in future discussions.}. 
The MHV vertices from the classical MHV Lagrangian
only generate the cut-constructible parts of YM loop amplitudes, such as the
one-loop MHV amplitude. These new contributions might be expected to lead to the missing,
rational parts. This would also potentially explain why in \cite{Cachazoetal0406}
the combination of all-plus vertices with MHV tree vertices did not yield the correct
single-minus amplitudes -- these additional MHV contributions
are missing.

Further evidence for the conjecture that the quantum MHV
Lagrangian is equivalent to quantum YM theory would
be welcome. One could start with seeking explicit proofs of the above proposals.
One can also investigate beyond massless one-loop gauge theory -- an advantage of the
Lagrangian approach is that the inclusion of masses, and of fermions and scalars, is
in principle clear.
There are other issues raised by this work. It is plausible that the potential
quantum versions of the twistor space formulations of
gauge theory \cite{Boelsetal06,Boelsetal07,Boels07} are most likely to be
allied to the quantum theory discussed here -- one simple reason for believing this
is that the regularisation employed here keeps one in four dimensions. Perhaps there are
simple twistor space analogues of the counterterms discussed above.

Finally, although for our purposes the lightcone worldsheet approach to
perturbative gauge theory provided simply the motivation for a particular
choice of regularisation scheme, we believe that it would be fruitful to
further explore possible connections between that framework and the
twistor string programme.

\noindent {\bf Addendum:} We would like to thank Paul Mansfield and Tim Morris
for having informed us that they have recently been pursuing research related to 
that presented in this paper. Their work, which is complementary to ours in that 
it employs dimensional regularisation, has now appeared in \cite{EFFMM07}.


                \section*{Acknowledgements}


It is a pleasure to thank Paul Heslop, Gregory Korchemsky, Paul Mansfield,
Tim Morris and Adele Nasti for discussions.
We would like to thank PPARC for support under the
Rolling Grant PP/D507323/1 and the Special Programme Grant PP/C50426X/1.
The work of GT is supported by an EPSRC Advanced Fellowship EP/C544242/1
and by an EPSRC Standard Research Grant EP/C544250/1.

\newpage

\appendix

\section{Notation}\label{notation}

\paragraph{Lightcone conventions}\mbox{}

Here we summarise our lightcone conventions. 
We start off  by introducing lightcone coordinates
\beq
 x^{\pm} :=  {x^0 \pm x^3  \over
\sqrt{2}} \ ,  \quad x^{z} :=  {x^1 + i  x^2  \over
\sqrt{2}} \ , \quad x^{\zb } :=  {x^1 - i x^2  \over
\sqrt{2}} \ . 
\eeq
We also have $x^+ = x_{-}$, $ x^{z} = -  x_{\zb}$,  and so on.  
The scalar product between two vectors $A$ and $B$ is written as 
\beq
A \cdot B := A_+ B_{-} +
A_{-} B_+ - A_{z} B_{\zb} - A_{\zb} B_{z}
\ . 
\eeq
We choose $x^-$ as our lightcone time coordinate, therefore
the lightcone gauge used in this paper is defined by
\beq
\label{lc} A^- = 0 \ .
\eeq
This condition can be written as $\eta \cdot A =0$, where $\eta$ is a constant null vector, chosen to have components  
$\eta := ( 1/\sqrt{2},0, 0,1/ \sqrt{2})$ (hence $\eta_{-} = 1$, $\eta_{+} = \eta_{z} = \eta_{\zb} = 0$). 

To any four-vector $p$  we associate the bispinor $p_{a \dot{a}}$ defined by 
\beq
\label{paad}
p_{a \dot{a}} \ := \ 
\sqrt{2}
\left( 
 \begin{matrix}
p_{-} & - p_z \\ - p_{\zb} & p_{+} 
\end{matrix}
\right)
\ . 
\eeq
We also define holomorphic and anti-holomorphic spinors as
\beq
\lambda_{a} \ : = \ {2^{1 \over 4} \over \sqrt{p_+}} 
\left( 
\begin{matrix} 
- p_{z}  \\ 
\, p_+ 
\end{matrix}
\right) 
\ , 
\qquad 
\lt_{\dot{a}} \ : = \ {2^{1 \over 4} \over \sqrt{p_+}} 
\left( 
\begin{matrix} 
- p_{\zb}  \\ 
\, p_+ 
\end{matrix}
\right) 
\ , 
\eeq
from which it follows that 
\beq
\label{paados}
\lambda_{a}\lt_{\dot{a}}  \ := \ 
\sqrt{2}
\left( 
 \begin{matrix}
{p_z p_{\zb} \over p_{+} }& - p_z \\ - p_{\zb} & p_{+} 
\end{matrix}
\right)
\ . 
\eeq
This is of course consistent with the on-shell condition $p_{-} = p_z p_{\zb} /  p_{+}$.  
Furthermore, comparing \eqref{paad} and \eqref{paados} and choosing $\eta$ as specified earlier,  
we see that a generic   off-shell vector $p$ can be decomposed as 
\beq
\label{mhvl1}
p \ = \ \lambda \lt \, + \, z \eta
\ , 
\eeq
where 
\beq
\label{mhvl2}
z \ = \ { p_{-} p_{+} - p_z p_{\zb} \over p_{+} \eta_{-}} \ = \ {p^2 \over 2 (p\cdot \eta )}
\ . 
\eeq
\eqref{mhvl1} and \eqref{mhvl2} are the familiar  decompositions of off-shell vectors 
in the MHV literature \cite{dk,bst04,bbst1,bbst2}.

The off-shell holomorphic spinor product is defined as:
\be \label{product}
\ip{ij}=\sqrt{2}\, \frac{p_+^ip_z^j-p_+^jp_z^i}{\sqrt{p_+^ip_+^j}}
\ , 
\ee
whereas for the antiholomorphic spinors we define
\be \label{product2}
[ij]=\sqrt{2}\, \frac{p_+^ip_{\zb}^j-p_+^jp_{\zb}^i}{\sqrt{p_+^ip_+^j}}
\ .  
\ee
In these conventions, one finds 
\beq
2 (p^i  \cdot p^j ) \ = \ \lan i  \, j  \ran \,[i\, j]  + \, \left( {p^j_{+} \over p^i_{+}}\right) (p^i)^2
\, + \, 
\left( {p^i_{+} \over p^j_{+}}\right) (p^j)^2  
\ , 
\eeq
or, in the case where $p^i$ and $p^j $ are on shell, $2 (p^i  \cdot p^j )  = \lan i  \, j  \ran \,[i\, j] $. 
In the standard QCD literature conventions it is customary to define   $2 (p^i  \cdot p^j )  =  \lan i  \, j  \ran \,[j\, i] $; 
this can be obtained by simply re-defining the inner product of two anti-holomorphic spinors, $[i\, j]$, to be the negative 
of the right hand side of \eqref{product2}. 

\paragraph{Useful identities}\mbox{}

The form (\ref{product}) is very convenient for
deriving identities for
$\ip{ij}$ that also involve the $p_+$ components. For instance,
one has:
\be
\begin{split}
& \sqrt{p_+^i}\ip{jk}+\sqrt{p_+^j}\ip{ki}+\sqrt{p_+^k}\ip{ij} \\
&= \sqrt{2}\, \frac{p_+^i(p_+^jp_z^k-p_+^kp_z^j)}{\sqrt{p_+^ip_+^jp_+^k}}
+\sqrt{2}\, \frac{p_+^j(p_+^kp_z^i-p_+^ip_z^k)}{\sqrt{p_+^ip_+^jp_+^k}}
+\sqrt{2}\, \frac{p_+^k(p_+^ip_z^j-p_+^jp_z^i)}{\sqrt{p_+^ip_+^jp_+^k}}=0
\ .
\end{split}
\ee
It is also easy to see how to apply momentum conservation,
take say $\ip{ij}$, and substitute
\be
p^j=-\sum_{k\neq j} p^k\;\;\quad \text{(for each component)}.
\ee
Then we have
\be
\sqrt{p_+^j}\ip{ij}=
\sqrt{2}\frac{p_+^i(-\sum_{k\neq j} p_z^k)+
(\sum_{k\neq j}p_+^k)p_z^i}{\sqrt{p_+^i}}=
-\sqrt{2}\sum_{k\neq j}\sqrt{p_+^k}\frac{p_+^ip_z^k-p_+^kp_z^i}
{\sqrt{p_+^ip_+^k}}
=\sum_{k\neq j}\sqrt{p_+^k}\ip{ki}
\ .
\ee
We have also used the momentum  bracket notation from \cite{EttleMorris}
\be
(ij) = p_+^ip_z^j-p_+^jp_z^i
\ , \quad \{ij\} = p_+^ip_{\bar z}^j-p_+^jp_{\bar z}^i
\ .
\ee

\paragraph{Lightcone Yang--Mills action}\mbox{}

Here we give the form of the lightcone Yang--Mills action that we use
in this paper. As discussed in more detail in \cite{BST06}, starting from
the YM Lagrangian $-(1/4)\, \tr F^2$, imposing the lightcone gauge (\ref{lc}), 
and integrating out the $A^+$ component which appears quadratically,
the final lightcone theory contains only the two physical components
$A_z$ and $A_\zb$ \cite{Tomboulis73,ScherkSchwarz75,Capper:1983hh},  which we associate with positive and negative helicity 
respectively. The Lagrangian takes the simple form (\ref{lcgeneral})
\be 
\Lcal_{\rm YM}=\Lcal_{+-}+\Lcal_{++-}+\Lcal_{--+}+\Lcal_{++--} \ ,
\ee
with
\be
\begin{split}
\Lcal_{+-}&=-2\, \tr\{ A_\zb(\p_+\p_--\p_z\p_\zb) A_z\} \ , 
\\
\Lcal_{++-}&=2ig\, \tr\{[A_z,\p_+A_\zb](\p_+)^{-1}(\p_\zb A_z)\} \ , 
\\
\Lcal_{--+}&=2ig\, \tr\{[A_\zb,\p_+A_z](\p_+)^{-1}(\p_z A_\zb)\} \ , 
\\
\Lcal_{++--}&=-2g^2\, \tr\{[A_\zb,\p_+A_z](\p_+)^{-2}[A_z,\p_+A_\zb]\}\ .
\end{split}
\ee
Note that, in agreement with CQT, we have used the normalisation
$\tr \{T^aT^b\}=\delta^{ab}$. In order to convert to the usual conventions for
Yang--Mills theory, we therefore need to rescale $g\ra g/\sqrt2$.

\paragraph{Relation to the notation of CQT}\mbox{}

 To compare our notation to that of \cite{Thorn05,CQT1,CQT2}, note that 
we employ outgoing momenta instead of incoming, therefore the all--plus
amplitudes in these works would be all--minus from our perspective, and
should thus be conjugated when comparing. Also, our time evolution
coordinate is taken to be $x^-$ rather that $x^+$, which (among other
changes) implies that $p^+$ of CQT becomes $p_+$. Our metric is also taken
to have opposite signature to that in CQT. Finally, CQT define momentum
brackets $K^\wedge_{ij}$ and $K^\vee_{ij}$, which are just our $(ij)$ and
$\{ij\}$ brackets respectively.

\section{Details on the four--point calculation} \label{appdetails}

In this appendix we prove two results that were used in section  \ref{allplus4section},
namely equations (\ref{ident1}) and (\ref{ident2}). To make the expressions more compact, 
instead of momentum brackets we use the following notation:
\be
f_{ij}=-\frac{(ij)}{p_+^ip_+^j}=\frac{p_z^i}{p_+^i}-\frac{p_z^j}{p_+^j}
\ .
\ee
The $f_{ij}$ variables satisfy the simple relation:
\be
f_{ij}=f_{ik}+f_{kj}
\ ,
\ee
while momentum conservation is applied as
\be
p_+^if_{ij}=-\sum p_+^k f_{kj}
\ .
\ee
Also, to minimise clutter, in this appendix we use the notation $q_i:=p_+^i$. 

\paragraph{Proof of the quadratic identity} \mbox{}

In order to show (\ref{ident1}), it is convenient to divide out by the 
$\sqrt{\pplfour}$ factor (which is there anyway in (\ref{initial})) in order 
to bring it to the form
\be \label{appident1}
\begin{split}
&q_4^2f_{34}f_{41}
+q_1^2f_{12}f_{41}
+q_2^2f_{12}f_{23}
+q_3^2f_{23}f_{34}\\
&-(q_2+q_3)(q_1+q_4)f_{12}f_{34}
-(q_3+q_4)(q_2+q_1)f_{23}f_{41}=0\ ,
\end{split}
\ee
Expanding out the two last terms in (\ref{appident1}) as
\be \label{partident2}
-(q_1q_3+q_2q_4)(f_{12}f_{34}+f_{23}f_{41})
-(q_1q_2+q_3q_4)f_{12}f_{34}-(q_2q_3+q_4q_1)f_{23}f_{41}\;,
\ee
we apply momentum conservation on each of the four components of 
the first term of (\ref{partident2}), in the following way:
\be
\begin{split}
&-q_1q_3f_{12}f_{34}=q_1f_{12}(q_1f_{14}+q_2f_{24})=-q_1^2f_{12}f_{41}+q_1q_2f_{12}f_{24}\;,\\
&-q_1q_3f_{23}f_{41}=q_3f_{23}(q_2f_{42}+q_3f_{43})=-q_3^2f_{23}f_{34}+q_2q_3f_{23}f_{42}\;,\\
&-q_2q_4f_{12}f_{34}=q_4(q_3f_{13}+q_4f_{14})f_{34}=-q_4^2f_{34}f_{41}+q_3q_4f_{13}f_{34}\;,\\
&-q_2q_4f_{23}f_{41}=q_2f_{23}(q_2f_{21}+q_3f_{31})=-q_2^2f_{12}f_{23}+q_2q_3f_{31}f_{23}\;.\\
\end{split}
\ee
Clearly these transformations have been chosen to cancel the first four terms in (\ref{appident1}).
Collecting the remaining terms, we obtain
\be
\begin{split}
&q_1q_2f_{12}(f_{24}-f_{34})+q_2q_3f_{23}(f_{42}+f_{31}-f_{41})+q_3q_4f_{34}(f_{13}-f_{12})-q_1q_4f_{23}f_{41}\\
&=q_1q_2f_{12}f_{23}+q_2q_3f_{23}f_{32}+q_3q_4f_{34}f_{23}+q_1q_4f_{23}f_{14}\\
&=f_{23}[q_2(q_1f_{12}+q_3f_{32})+q_4(q_3f_{34}+q_1f_{14})]=f_{23}[-q_2(q_4f_{42})-q_4(q_2f_{24})]\\
&=0
\end{split}
\ee
thus showing (\ref{ident1}).

\paragraph{Proof of the linear identity}\mbox{}

We will now outline the proof ot the linear (in region momenta) identity
(\ref{ident2}). Converting it to the notation used in the appendix, and
performing simple manipulations, we find (suppressing the overall $3/8$ factor):
\be
\begin{split}
X=\ &q_4^2((\bp_3+\bp_4)+(\bp_2+\bp_3))f_{34}f_{41}
+q_1^2(-(\bp_2+\bp_3)+(\bp_3+\bp_4))f_{12}f_{41}\\
&+q_2^2(-(\bp_3+\bp_4)-(\bp_2+\bp_3))f_{12}f_{23}
+q_3^2(+(\bp_2+\bp_3)-(\bp_3+\bp_4))f_{23}f_{34}\\
&-\half(q_2+q_3)(q_1+q_4)[(\bp_3-\bp_2)+(\bp_1-\bp_4)]f_{12}f_{34}\\
&-\half(q_3+q_4)(q_1+q_2)[(\bp_4-\bp_3)+(\bp_2-\bp_1)]f_{23}f_{41}\\
=\ &(\bp_3-\bp_1)(q_4^2f_{34}f_{41}-q_2^2f_{12}f_{23})
+(\bp_4-\bp_2)(q_1^2f_{12}f_{41}-q_3^2f_{23}f_{34})\\
&-(q_2+q_3)(q_1+q_4)(\bp_3+\bp_1)f_{12}f_{34}
-(q_3+q_4)(q_1+q_2)(\bp_2+\bp_4)f_{23}f_{41}\\
=\ &(\bp_3-\bp_1)(q_4^2f_{34}f_{41}-q_2^2f_{12}f_{23})
+(\bp_4-\bp_2)(q_1^2f_{12}f_{41}-q_3^2f_{23}f_{34})\\
&-(\bp_1+\bp_3)q_2q_4(f_{12}f_{34}-f_{23}f_{41})+(\bp_2+\bp_4)q_1q_3(f_{12}f_{34}-f_{23}f_{41})\\
&-(\bp_1+\bp_3)(q_1q_2+q_3q_4)f_{12}f_{34}+(\bp_1+\bp_3)(q_2q_3+q_4q_1)f_{23}f_{41}\ .
\end{split}
\ee
Similarly to the previous case, we will rewrite the second line in the final expression
in such a way that we completely cancel all the terms in the first line.
To do that we use
\be
\begin{split}
-(\bp_1+\bp_3)q_2q_4(f_{12}f_{34}-f_{23}f_{41})=&
(\bp_3-\bp_1)(q_2^2f_{12}f_{23}-q_4^2f_{34}f_{41})+\\
&+q_1q_2\bp^1f_{12}f_{31}-q_4q_1\bp^1f_{41}f_{13}+\\
&+q_3q_4\bp^3f_{34}f_{13}-q_2q_3\bp^3f_{23}f_{31}
\end{split}
\ee
and
\be
\begin{split}
(\bp_2+\bp_4)q_1q_3(f_{12}f_{34}-f_{23}f_{41})=&
(\bp_4-\bp_2)(q_3^2f_{23}f_{34}-q_1^2f_{12}f_{41})+\\
&+q_2q_3\bp_2f_{23}f_{42}-q_1q_2\bp_2f_{12}f_{24}+\\
&+q_4q_1\bp_4f_{41}f_{24}-q_3q_4\bp_4f_{34}f_{42}\ .
\end{split}
\ee
What remains after substituting these is
\be
\begin{split}
X=\ &\bp_1q_1f_{31}(q_2f_{12}+q_4f_{41})+q_3\bp_3f_{13}(q_4f_{34}+q_2f_{23})\\
&+\bp_2q_2f_{42}(q_3f_{23}+q_1f_{12})+q_4\bp_4f_{24}(q_1f_{41}+q_3f_{34})\\
&-(\bp_1+\bp_3)(q_1q_2+q_3q_4)f_{12}f_{34}+(\bp_1+\bp_3)(q_2q_3+q_4q_1)f_{23}f_{41}\\
=\ &\bp_1q_1q_2f_{12}f_{41}+\bp_3q_3q_4f_{34}f_{23}
+\bp_1q_4q_1f_{41}f_{21}+\bp_3q_2q_3f_{23}f_{43}\\
&+\bp_2q_2f_{42}(q_3f_{23}+q_1f_{12})+q_4\bp_4f_{24}(q_1f_{41}+q_3f_{34})\\
&-(\bp_1q_3q_4+\bp_3q_1q_2)f_{12}f_{34}+(\bp_1q_2q_3+\bp_3q_4q_1)f_{23}f_{41}\ .
\end{split}
\ee
Now we collect various terms together to rewrite $X$ as
\be
\begin{split}
X=\ &\bp_1q_2f_{41}(q_1f_{12}+q_3f_{23})+\bp_3q_4f_{23}(q_3f_{34}+q_1f_{41})\\
&+\bp_1q_4f_{21}(q_1f_{41}+q_3f_{34})+\bp_3q_2f_{43}(q_3f_{23}+q_1f_{12})\\
&+\bp_2q_2f_{42}(q_3f_{23}+q_1f_{12})+\bp_4q_4f_{24}(q_1f_{41}+q_3f_{34})\\
=\ &\bp_1q_2f_{41}(2q_3f_{23}-q_4f_{42})+\bp_3q_4f_{23}(2q_1f_{41}-q_2f_{24})\\
&+\bp_1q_4f_{21}(2q_1f_{41}-q_4f_{42})+\bp_3q_2f_{43}(2q_3f_{23}-q_4f_{42})\\
&+\bp_2q_2f_{42}(2q_3f_{23}-q_4f_{42})+\bp_4q_4f_{24}(2q_1f_{41}-q_2f_{24})\\
=\ &2[q_2q_3f_{23}(\bp_1f_{41}+\bp_3f_{43}+\bp_2f_{42})
+q_4q_1f_{41}(\bp_3f_{23}+\bp_1f_{21}+\bp_4f_{24})] \\
&+(\bp_1+\bp_2+\bp_3+\bp_4)q_2q_4f_{24}f_{42}\ .
\end{split}
\ee
Clearly the term on the last line vanishes by momentum conservation.
We now restore all labels to write the final result as
\be
\begin{split}
X=&2~(32)[f_4(p_\zb^1+p_\zb^2+p_\zb^3)-p_\zb^1f_1-p_\zb^2f_2-p_\zb^3f_3]+\\
&+2~(14)[f_2(p_\zb^1+p_\zb^3+p_\zb^4)-p_\zb^3f_3-p_\zb^1f_1-p_\zb^4f_4]\ ,
\end{split}
\ee
where we used that $q_2q_3f_{23}=p_+^2p_+^3(p_z^2/q_+^2-p_z^3/p_+^3)
=p_+^3p_z^2-p_+^2p_z^3=(32)$ (and similarly for $(14)$), and where $f_i=p_z^i/p_+^i$.
Using momentum conservation on both terms, we rewrite them as
\be
X=-2[(32)+(14)]\left[\frac{p_\zb^1p_z^1}{p_+^1}+\frac{p_\zb^2p_z^2}{p_+^2}+
\frac{p_\zb^3p_z^3}{p_+^3}+\frac{p_\zb^4p_z^4}{p_+^4}\right]
\ .
\ee
For each momentum we have that $p^2=2(p_+p_--p_zp_\zb)$, therefore  we can rewrite the above as
\be
X=+[(32)+(14)]\left[\frac{(p_1)^2}{p_+^1}+\frac{(p_2)^2}{p_+^2}+
\frac{(p_3)^2}{p_+^3}+\frac{(p_4)^2}{p_+^4}
+2(p_-^1+p_-^2+p_-^3+p_-^4)\right]
\ .
\ee
The $p_-$ term vanishes,   hence, noticing also that
$(32)+(14)=-\half((12)+(23)+(34)+(41))$, we conclude that
\be
X=-\half[(12)+(23)+(34)+(41)]\sum_{i=1}^4 \frac{(p_i)^2}{p_+^i}
\ .
\ee

\paragraph{Off-shell terms in the four-point case} \mbox{}

For completeness, we also give the form of the off-shell terms that arose in 
the manipulations leading to \eqref{finalfinal}.

Using the notation $P_{ij}=(\frac{(p_i)^2}{p_+^i}+\frac{(p_j)^2}{p_+^j})$
they are :
\be \label{simp}
\begin{split}
&f(p^2)=\frac{1}{4\ip{12}\cdots\ip{41}}\big[-P_{13}(\bp_1+\bp_2)(41)-P_{13}(\bp_2+\bp_3)(12)+P_{24}(\bp_2+\bp_3)(42)\\
&+\frac{1}{p_+^1}P_{12} [(p_+^2+p_+^3)(2\bp_1+\bp_2-\bp_3)-p_+^3(\bp_1+\bp_2)-p_+^1(\bp_2+\bp_3)]~ (13)\\
&+P_{12}\frac{p_+^3}{p_+^1p_+^2}[p_+^2(\bp_1+\bp_2)-p_+^4(\bp_2+\bp_3)](12)
-2P_{13}\frac{1}{p_+^1}\{31\}(41)\big] \;.
\end{split}
\ee
This expression, together with ${\cal V}^{(4)}_k$ in \eqref{regionvertex}, 
should be added to \eqref{finalfinal} in order to recover a fully off-shell 
four--point vertex. 

\newpage

\bibliography{ctrefs}
\bibliographystyle{JHEP1}

\end{document}